%% file: main.tex
\documentclass[9pt,twocolumn,twoside]{pnas-new}

\templatetype{pnasresearcharticle}

\usepackage{lipsum}
\usepackage{makecell}
\usepackage{float}
\floatstyle{plaintop}
\newfloat{algorithm}{t}{lop}
\floatname{algorithm}{Algorithm}

\begin{document}


\title{Design of 3D Non-Cartesian Trajectories for Fast Volumetric MRI via Analytic Coordinate Discretization}

\author[a,1]{\orcid{0009-0000-7031-1347}{Kwang Eun~Jang}}
\author[a]{\orcid{0009-0005-5963-765X}{Dwight G.~Nishimura}}

\affil[a]{Magnetic Resonance Systems Research Lab (MRSRL), Department of Electrical Engineering, Stanford University}

\leadauthor{Jang}

\significancestatement{
    3D non-Cartesian trajectories offer several advantages over rectilinear trajectories for rapid volumetric imaging, including improved sampling efficiency and greater robustness to motion, flow, and aliasing artifacts. However, the design of 3D non-Cartesian trajectories is nontrivial and often involve complicated process. This work introduces an analytical framework that simplifies both trajectory design and density compensation, leading to more efficient and flexible sampling for 3D MRI with shorter scan times and fewer artifacts.
}

\correspondingauthor{\textsuperscript{1}To whom correspondence should be addressed. E-mail: kejang@stanford.edu, kejang@gmail.com}


\begin{abstract}
    3D non-Cartesian trajectories offer several advantages over rectilinear trajectories for rapid volumetric imaging, including improved sampling efficiency and greater robustness to motion, flow, and aliasing artifacts. In this paper, we present a unified framework for designing three widely used non-Cartesian trajectories: 3D Radial, 3D Cones, and Stack-of-Spirals. Our approach is based on the idea that a non-Cartesian trajectory can be interpreted as a discretized version of an analytic coordinate defined by a set of template trajectories. Equivalently, the analytic coordinate is conceptualized as a non-Cartesian trajectory composed of an infinite number of copies of a set of template trajectories. The discretization is accomplished by constructing a continuous spiral path on a surface and sampling points along this path at unit intervals, leaving only the essential spokes/interleaves, thereby yielding the practical non-Cartesian trajectory from the analytic coordinate. One of the advantages of our approach is that the analytic density compensation factor can be readily derived using Jacobian determinants, which quantify changes in unit areas due to the transformation from the analytic coordinate to the Cartesian grid. Additionally, the proposed approach derives analytic formulae to compute the number of readouts based on prescribed parameters, allowing us to specify the trajectory's acceleration factor for a given total scan time. Furthermore, variable-density sampling can be easily incorporated, and spokes/interleaves are smoothly distributed in k-space along the derived spiral path, even for a small number of readouts. In a preliminary phantom study, the proposed method demonstrated improved sampling efficiency and image quality compared to the conventional approach.
\end{abstract}

\dates{This manuscript was compiled on \today}

\maketitle
\thispagestyle{firststyle}
\ifthenelse{\boolean{shortarticle}}{\ifthenelse{\boolean{singlecolumn}}{\abscontentformatted}{\abscontent}}{}

\firstpage[4]{5}


\input{sections/intro.tex}

\input{sections/prelim.tex}

\input{sections/radial.tex}

\input{sections/cones.tex}

\input{sections/spiral.tex}

\input{sections/result.tex}

\input{sections/discussion.tex}

\input{sections/acknowledgement.tex}

\input{sections/diffeq.tex}

\input{sections/simul-radial-isotropic.tex}

\input{sections/simul-radial-anisotropic.tex}

\end{document}

%% file: sections/intro.tex

\dropcap{A}t each instant, MR scanners acquire data at a single point in k-space. This point travels along a trajectory during MR scans to sufficiently populate multidimensional k-space. This trajectory determines various imaging characteristics, such as scan time, spatial resolution, field-of-view (FOV), aliasing artifacts, image contrast, and robustness or susceptibility to certain imperfections.

Rectilinear rasters, also referred to as Cartesian trajectories, have been predominantly used in MR imaging \cite{bernstein2004handbook}. Cartesian trajectories are immune to many system imperfections, such as B0 inhomogeneity, eddy currents, and gradient delays \cite{bernstein2004handbook}. Additionally, reconstruction with Cartesian trajectories is relatively straightforward, often requiring only a single execution of the inverse fast Fourier transform \cite{bernstein2004handbook}.

However, Cartesian trajectories have several drawbacks. Data are acquired only after phase encoding steps, leading to delayed echo times and prolonged sequence durations. Furthermore, Cartesian trajectories sweep the entire cuboid from corner to corner, even though filling an ellipsoid is often sufficient, further extending scan time. Additionally, Cartesian trajectories are prone to motion and flow artifacts \cite{nishimura_velocity_1995, irarrazabal_fast_1995} since the center of k-space is sampled only a few times.

3D non-Cartesian trajectories have been primarily studied for rapid volumetric imaging \cite{irarrazabal_fast_1995, larson_anisotropic_2008, gurney_design_2006, pipe_new_2011, piccini_spiral_2011, turley_distributed_2013, feng_xd-grasp_2016, feng_golden-angle_2022}, especially in cardiac applications \cite{thedens_fast_1999, addy_high-resolution_2015, usman_free_2017, malave_whole-heart_2019}. Many 3D non-Cartesian trajectories fill an ellipsoid rather than a cuboid, improving sampling efficiency. Additionally, non-Cartesian trajectories are often designed without phase encoding steps, initiating data acquisition as soon as the trajectory begins at the center of k-space, which reduces echo times and sequence repetition times. Due to repeated acquisitions of the center of k-space, they are also robust against motion and flow artifacts \cite{bernstein2004handbook, nishimura_velocity_1995, irarrazabal_fast_1995, larson_anisotropic_2008, piccini_spiral_2011, feng_golden-angle_2022}. Moreover, this dense sampling around the origin reduces aliasing artifacts, enabling more aggressive undersampling.

In this paper, we introduce a unified method for designing three widely used non-Cartesian trajectories: 3D Radial, 3D Cones, and Stack-of-Spirals. The fundamental idea behind our approach is that trajectories can be seen as discretized versions of their respective analytic coordinates. Take, for instance, 2DFT, the most widely used Cartesian trajectory, which consists of multiple horizontal lines spaced by $\Delta k_y = 1/\text{FOV}_y$, with each line comprised of samples spaced by $\Delta k_x = 1/{BW}$. Therefore, we view 2DFT as a discretized form of the 2D Cartesian coordinate, fine-tuned to specific imaging parameters. In a similar vein, we structure the design of non-Cartesian trajectories as outlined below:
\begin{enumerate}
    \item Define an analytic coordinate. This coordinate is conceptualized as a non-Cartesian trajectory comprised of infinite copies of a set of template trajectories. For 3D Radial, 3D Cones, and stack-of-spirals trajectories, these templates correspond to 3D spokes, conic interleaves, and spiral interleaves, respectively.
    \item Discretize the analytic coordinate by sketching a spiral path on a surface and subsequently discretizing this path at unit intervals.
    \item Determine the density compensation factor (DCF). The DCF is proportional to both the Jacobian determinant \cite{hoge_density_1997} associated with the transition from the analytic non-Cartesian coordinate to the Cartesian coordinate and the reciprocal of the distribution of the discretized samples. The latter is inferred from the spiral path.
\end{enumerate}

Although our proposed design methodology was originally developed for 3D Cones trajectories \cite{jang2017redesigned, jang2018redesigned}, we begin by introducing the design of simple center-out 3D Radial trajectories. These trajectories are conceptually and computationally straightforward, making them easier to derive and verify. We then extend the methodology to the design of 3D Cones and Stack-of-Spirals trajectories by discretizing their respective analytic coordinates.

%% file: sections/prelim.tex
\section*{Preliminaries}

Table~\ref{tab:notation} summarizes the notations used throughout this paper. In our non-Cartesian trajectory design methods, this differential equation appears:
\begin{equation}
    \frac{df}{du} = \dot{f}(u) = \frac{N}{g(f(u))},\label{eq:diffeq_problem}
\end{equation}
where $u$ is a bounded variable ranging from 0 to 1, $g(\cdot)$ is a known function, $N$ denotes a constant, and $f(u)$ is a function with a positive, monotonically increasing inverse, $u(f)$. We also assume that $f(u)$ is bounded such that $f(u) \in (f_{\min}, f_{\max})$. Algorithm~\ref{alg:diffeq} provides a numerical solution to \eqref{eq:diffeq_problem}, with its derivation presented in Appendix 1.

\begin{table}[t!]
    \centering
    \caption{Notations}\label{tab:notation}
    \begin{tabular}{ll}
    \toprule
    \textbf{Symbol} & \textbf{Definition} \\
    \midrule
    $\vec{k}(t)$ & Sampling location in k-space at time $t$ \\
    $k(t)$ & Distance from the origin: \\
           & $\makecell{ k(t) = \|\vec{k}(t)\| = \sqrt{k_x^2(t) + k_y^2(t) + k_z^2(t)} }$ \\[3pt]
    $\vec{g}(t)$ & Gradient waveform: \\
                 & $\makecell{\vec{g}(t) = [g_x(t), g_y(t), g_z(t)]^T}$ \\[3pt]
    $g(t)$ & Gradient amplitude: \\
           & $\makecell{ g(t) = \|\vec{g}(t)\| = \sqrt{g_x^2(t) + g_y^2(t) + g_z^2(t)} }$ \\[3pt]
    $s(t)$ & Slew-rate: \\
           & $\makecell{ s(t) = \sqrt{\left({d g_x}/{dt}\right)^2 + \left({d g_y}/{dt}\right)^2 + \left({d g_z}/{dt}\right)^2}}$ \\[3pt]
    $\vec{k}_r(t)$ & Radial k-space location: \\
                   & $\makecell{\vec{k}_r(t) = [k_x(t), k_y(t)]^T}$ \\[3pt]
    $k_r(t)$ & Radial k-space distance: \\
             & $\makecell{ k_r(t) = \|\vec{k}_r(t)\| = \sqrt{k_x^2(t) + k_y^2(t)}}$ \\[3pt]
    $\vec{g}_r(t)$ & Radial gradient: \\
                   & $\makecell{\vec{g}_r(t) = [g_x(t), g_y(t)]^T}$ \\[3pt]
    $g_r(t)$ & Radial gradient amplitude: \\
             & $\makecell{ g_r(t) = \|\vec{g}_r(t)\| = \sqrt{g_x^2(t) + g_y^2(t)}}$ \\[3pt]
       $K_r$ & Radial extent in k-space: \\
             & $\makecell{K_r = 1/(2 \Delta x) = 1/(2 \Delta y)}$ \\[3pt]
       $K_z$ & Axial extent in k-space: \\
             & $\makecell{K_z = 1/(2 \Delta z)}$ \\[3pt]
    $K(\phi)$ & Extent at polar angle $\phi$ in k-space: \\
                     & $\makecell{K(\phi) = {K_r K_z}\big/{\sqrt{K_r^2 \cos^2 \phi + K_z^2 \sin^2 \phi}}.}$ \\[3pt]
    $L_r$, $L_z$ & FOVs in radial and z-directions\\
    $L(\phi)$ & Field-of-view (FOV) at polar angle $\phi$: \\
              & $\makecell{L(\phi) = {L_r L_z}\big/{\sqrt{L_r^2 \cos^2 \phi + L_z^2 \sin^2 \phi}}}$ \\[3pt]
    $L_{90^\circ}(\phi)$ & Orthogonal FOV: $L_{90^\circ}(\phi) = L(\phi + \pi / 2)$\\
    \bottomrule
    \end{tabular}
    \label{tab:notations}
\end{table}

\begin{algorithm}[t!]
    \centering
    \caption{Numerical Solution of \eqref{eq:diffeq_problem}}
    \begin{tabular}{p{0.05\textwidth}p{0.38\textwidth}}
    \toprule
    \textbf{Input} & Given function $g(\cdot)$ and the range of $f(\cdot)$. \\
    \textbf{Output} & $f(u)$ computed based on the provided function $g(\cdot)$. \\
    \midrule
    \textbf{Step 1} & Calculate $N$ by integrating $g(\cdot)$: 
    \[
    N = \int_{f_{\min}}^{f_{\max}} g(x) dx.
    \] \\
    \textbf{Step 2} & Determine $u(f)$ using the cumulative summation:
    \[
    u(f) = \frac{1}{N} \int_{f_{\min}}^{f} g(\tau) d\tau.
    \] \\
    \textbf{Step 3} & Compute the inverse of $u(f)$ over $u \in (0, 1)$. \\
    \bottomrule
    \end{tabular}
    \label{alg:diffeq}
\end{algorithm}

%% file: sections/radial.tex
\section*{3D Center-out Radial Trajectories}

\subsection*{3D Radial Coordinates}\label{subsec:traj-3dpr-coord}

Imagine a trajectory in k-space that begins at the origin and extends along the $k_z$-axis, which is often referred to as a ``spoke.'' This template spoke allows for the unique representation of any point within a certain radius by following these steps:
\begin{enumerate}
    \item Consider a point in a Spherical coordinate, represented as $(\rho, \theta, \phi)$, and a template spoke, $\vec{k}^{(0)}(t)$.
    \item Identify the value of $t$ for which $k^{(0)}(t) = \rho$.
    \item Rotate the template around the $k_y$-axis by the polar angle $\phi$.
    \item Further rotate the template around the $k_z$-axis by the azimuthal angle $\theta$.
\end{enumerate}

With this process, we introduce a new coordinate system, 3D Radial coordinate, which uniquely determines the location of a point using three coordinates, $(t, \theta_p, \phi_p)$, and the template, $\vec{k}^{(0)}(t)$:
\begin{equation}
    \begin{bmatrix}
        \rho\\
        \theta\\
        \phi
    \end{bmatrix} = 
    \begin{bmatrix}
        k^{(0)}(t)\\
        \theta_p\\
        \phi_p
    \end{bmatrix}.
\end{equation}

The 3D Radial coordinate can be envisioned as a 3D Radial trajectory composed of an infinite array of spokes. We design 3D Radial trajectories by discretizing this analytical 3D Radial coordinate with optimal spacing to ensure that only the essential spokes remain.

\subsection*{Discretization of the 3D Radial Coordinate}\label{subsec:traj-3dpr-discretization}

The time variable $t$ is inherently discretized when setting a readout bandwidth. The subsequent challenge is discretizing the azimuthal and polar angles, $(\theta_p, \phi_p)$, which is equivalent to sampling points on an ellipsoidal surface. One solution is to draw a spiral path on the ellipsoid and sample along this path. Consider a parametric spiral path on an ellipsoid\footnote{This ellipsoid is not necessarily in k-space or image-space, as the primary objective of this procedure is to acquire rotation angles rather than to directly construct trajectories in k-space. Here, we slightly abuse the notation of $x$, $y$, and $z$ to represent the 3D coordinate in this conceptual space.}:
\begin{equation}
    \begin{bmatrix}
        p_x(u) \\
        p_y(u) \\
        p_z(u)
    \end{bmatrix}
    =
    \begin{bmatrix}
        k_{\max, r} \cos\theta_p(u) \sin\phi_p(u) \\
        k_{\max, r} \sin\theta_p(u) \sin\phi_p(u) \\
        k_{\max, z} \cos\phi_p(u)
    \end{bmatrix},
    \label{eq:3dpr_spiral_path}
\end{equation}
where $u \in (0, 1)$. While $u$ is not directly related to any physical quantity, we interprete it as time on this path in the following derivation. The derivatives of $\theta_p(u)$ and $\phi_p(u)$ with respect to time $u$ represent the speeds of the spiral path in the azimuthal and polar directions, respectively. The objective is to determine $\theta_p(u)$ and $\phi_p(u)$ based on imaging parameters such as FOV, resolution, and the number of readouts. The discretized $\theta_p[u]$ and $\phi_p[u]$ are then acquired by uniformly sampling the parameter $u$ at intervals of $\Delta u$, where $\Delta u$ is one over the total number of readouts:
\begin{equation}
    \Delta u = \frac{1}{N}.
\end{equation} 

One complete rotation along a latitude circle takes:
\begin{equation}
    \frac{2 \pi}{\dot{\theta_p}(u)}.\label{eq:3dpr_duration}
\end{equation}

\begin{figure}[!htb]
    \centering
    \includegraphics[width=0.47\textwidth, keepaspectratio]{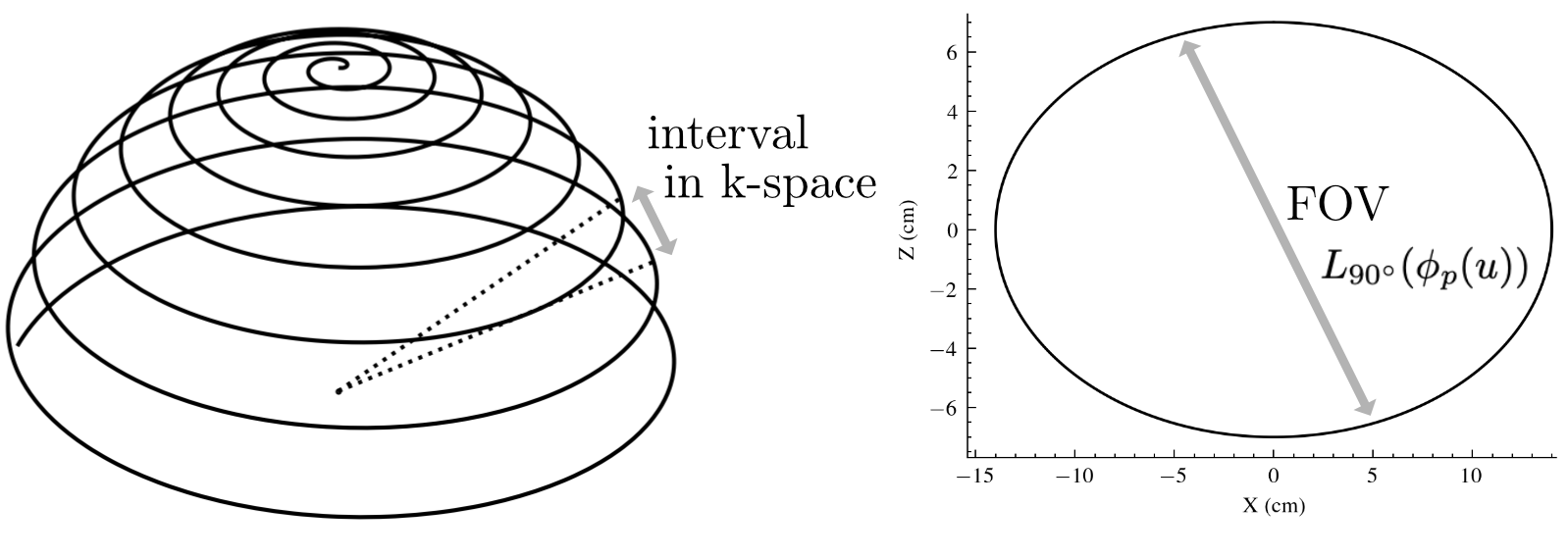}
    \caption[Relationship between the interval on the spiral path and the FOV in image-space.]{The distance between two adjacent latitude circles can be approximated using the length of an arc on the ellipsoid. This spacing determines the FOV in the orthogonal direction in image-space \cite{larson_anisotropic_2008}.}
    \label{fig:3dpr_orthogonal_fov}
\end{figure}

The distance between consecutive latitude circles can be approximated using the length of an arc on the ellipsoid. This distance should be the inverse of the FOV in the orthogonal direction \cite{larson_anisotropic_2008}, as illustrated in Figure \ref{fig:3dpr_orthogonal_fov}:
\begin{equation}
    \underbrace{
        \frac{2 \pi}{\dot{\theta_p}(u)}
    }_{\text{duration}}
    \underbrace{
        \vphantom{\frac{2 \pi}{\dot{\theta_p}(u)}}
        \dot{\phi}_p(u)
    }_{\text{speed}}
    \underbrace{
        \vphantom{\frac{2 \pi}{\dot{\theta_p}(u)}}
        K(\phi_p(u))
    }_{\text{radius}}
    = \frac{1}{L_{90 ^{\circ}}(\phi_p(u))}.
    \label{eq:3dpr_dist_along_polar}
\end{equation}

We compute the number of samples taken over the duration given in \eqref{eq:3dpr_duration} from both a temporal and a spatial perspective. From a temporal perspective, the number of samples is the duration divided by the sampling interval $\Delta u$: 
\begin{equation}
    N_{\text{samp}} = \underbrace{
        \left(\frac{2 \pi}{\dot{\theta_p}(u)}\right)
    }_{\text{duration}}
    \underbrace{
        \vphantom{\left(\frac{2 \pi}{\dot{\theta_p}(u)}\right)}
        (\Delta u)
    }_{\text{interval}}
    \vphantom{(\Delta u)}^{-1}.
    \label{eq:3dpr_along_latitude_1}
\end{equation}
From a spatial perspective, the number of samples should equal the circumference divided by the desired interval in k-space:
\begin{equation}
    N_{\text{samp}} = \underbrace{
        \vphantom{\left(\frac{1}{L_r}\right)}
        \left(2 \pi K(\phi_p(u)) \sin \phi_p(u)\right)
    }_{\text{circumference}}
    \underbrace{
        \left(\frac{1}{L_r}\right)
    }_{\text{interval}}
    \vphantom{\left(\frac{1}{L_r}\right)}^{-1}.\label{eq:3dpr_along_latitude_2}
\end{equation}

By setting \eqref{eq:3dpr_along_latitude_1} equal to \eqref{eq:3dpr_along_latitude_2}, we derive a differential equation for $\theta_p(u)$:
\begin{equation}
    \frac{2 \pi N}{\dot{\theta_p}(u)} = 2 \pi L_r K(\phi_p(u)) \sin \phi_p(u),
    \label{eq:3dpr_along_latitude}
\end{equation}
where $N = (\Delta u)^{-1}$ denotes the number of readouts. This is due to the uniform sampling over $u \in (0, 1)$.

From \eqref{eq:3dpr_dist_along_polar} and \eqref{eq:3dpr_along_latitude}, we derive the following differential equations for $\phi_p(u)$ and $\theta_p(u)$:
\begin{align}
    \dot{\phi}_p(u) &= \frac{N}{2 \pi L_r L_{90 ^{\circ}}(\phi_p(u)) K^2(\phi_p(u)) \sin \phi_p(u)},\label{eq:3dpr_diffeq_polar}\\
    \dot{\theta}_p(u) &= \frac{N}{L_r K(\phi_p(u))\sin\phi_p(u)}.\label{eq:3dpr_diffeq_theta}
\end{align}

Since \eqref{eq:3dpr_diffeq_polar} is a specific case of \eqref{eq:diffeq_problem}, we solve this problem using Algorithm \ref{alg:diffeq}. The elliptical models described in Table~\ref{tab:notation} are employed for anisotropic resolution and/or FOV. This procedure is summarized in Algorithm \ref{alg:3dpr_full}. 

\begin{algorithm}[t!]
    \centering
    \caption{Spiral Path on Ellipsoid for 3D Radial Trajectories}\label{alg:3dpr_full}
    \begin{tabular}{p{0.05\textwidth}p{0.38\textwidth}}
    \toprule
    \textbf{Input} & $L(\phi)$ and $K(\phi)$. \\
    \textbf{Output} & $\phi_p(u)$ and $\theta_p(u)$. \\
    \midrule
    \textbf{Step 1} & Calculate the number of readouts $N$: 
    \[
    N = 2 \pi L_r \int_0^\pi L_{90 ^{\circ}}(\phi) K^2(\phi) \sin \phi ~ d\phi. \label{eq:3dpr_nrd}
    \] \\
    \textbf{Step 2} & Determine $u(\phi)$ using cumulative summation:
    \[
    u(\phi) = \frac{1}{N} \int_{0}^{\phi} 2 \pi L_r L_{90 ^{\circ}}(\tau) K^2(\tau) \sin \tau d\tau. \label{eq:3dpr_u_of_phi}
    \] \\
    \textbf{Step 3} & $\phi_p(u)$: the inverse of $u(\phi)$ over $u \in (0, 1)$. \\
    \textbf{Step 4} & Determine $\theta_p(u)$ using the cumulative summation of:
    \[
    \dot{\theta}_p(u) = \frac{N}{L_r K(\phi_p(u))\sin\phi_p(u)}.
    \] \\[-8pt]
    \bottomrule
    \end{tabular}
\end{algorithm}

In Appendix 2, we compare the proposed method for designing a simple 3D Radial trajectory with the intuitive approach, where the spokes are constructed by connecting the origin to the tips of the spokes. Both approaches result in the same number of readouts when an isotropic FOV and isotropic resolution are prescribed.


\subsection*{Density Compensation Factor}\label{subsec:traj-3dpr-dcf}

An analytic density compensation factor (DCF) is derived using a method similar to Hoge et al. \cite{hoge_density_1997}. This derivation involves both the Jacobian determinant associated with the transformation of the 3D Radial coordinate to the Cartesian coordinate \cite{hoge_density_1997}, $\det(J_{\text{PR-to-Cart}})$, and the distribution of discretized samples.

To compute $\det(J_{\text{PR-to-Cart}})$, we first evaluate the Jacobian determinant between the 3D Radial coordinate and the Spherical coordinate:
\begin{align}
    \det(J_{\text{PR-to-Sph}}) &= 
    \begin{vmatrix}  
        \dfrac{\partial \rho}{\partial t} & \dfrac{\partial \theta}{\partial t} & \dfrac{\partial \phi}{\partial t}\nonumber\\[10pt]
        \dfrac{\partial \rho}{\partial \theta_p} & \dfrac{\partial \theta}{\partial \theta_p} & \dfrac{\partial \phi}{\partial \theta_p}\nonumber\\[10pt]
        \dfrac{\partial \rho}{\partial \phi_p} & \dfrac{\partial \theta}{\partial \phi_p} & \dfrac{\partial \phi}{\partial \phi_p}
    \end{vmatrix}\nonumber\\[10pt]
    &= \frac{\partial \rho}{\partial t} = \frac{\gamma}{2 \pi} \frac{(\vec{g} \cdot \vec{k})[t]}{k[t]}.
\end{align}
\noindent The Jacobian determinant between the Spherical coordinate and the Cartesian coordinate is:
\begin{equation}
    \det(J_{\text{Sph-to-Cart}}) = \rho^2 \sin \phi = k^2[t] \sin \phi_p[u].
\end{equation}

\noindent Consequently, the Jacobian determinant associated with the transformation of the 3D Radial coordinate to the Cartesian coordinate is:
\begin{equation}
    \det(J_{\text{PR-to-Cart}}) = \frac{\gamma}{2 \pi} k[t] (\vec{g} \cdot \vec{k})[t] \sin \phi_p[u].
\end{equation}

The denominator of $\dot{\phi}_p(u)$, represented by \(g(\cdot)\) in \eqref{alg:diffeq}, is proportional to the probability density function of discretized samples as shown in \eqref{eq:p_and_g}. Thus, the final DCF is computed as:
\begin{align}
    d[t, u] &\propto \frac{|\det(J_{\text{PR-to-Cart}})|}{g[u]}\nonumber\\
    &= \frac{|\det(J_{\text{PR-to-Cart}})|}{2 \pi L_r L_{90 ^{\circ}}(\phi_p(u)) K^2(\phi_p(u)) \sin \phi_p(u)}\label{eq:3dpr-dcf}\\
    &= \frac{\gamma}{4 \pi^2} \frac{k[t]|(\vec{g}\cdot\vec{k})[t]|}{k^2_{\max}(\phi_p[u]) L_{90^{\circ}}(\phi_p[u])}.\nonumber
\end{align}

%% file: sections/cones.tex
\section*{3D Cones Trajectories}

\begin{figure}[!htb]
    \centering
    \begin{minipage}{0.18\textwidth}
        \centering
        \includegraphics[height=3.2cm, keepaspectratio]{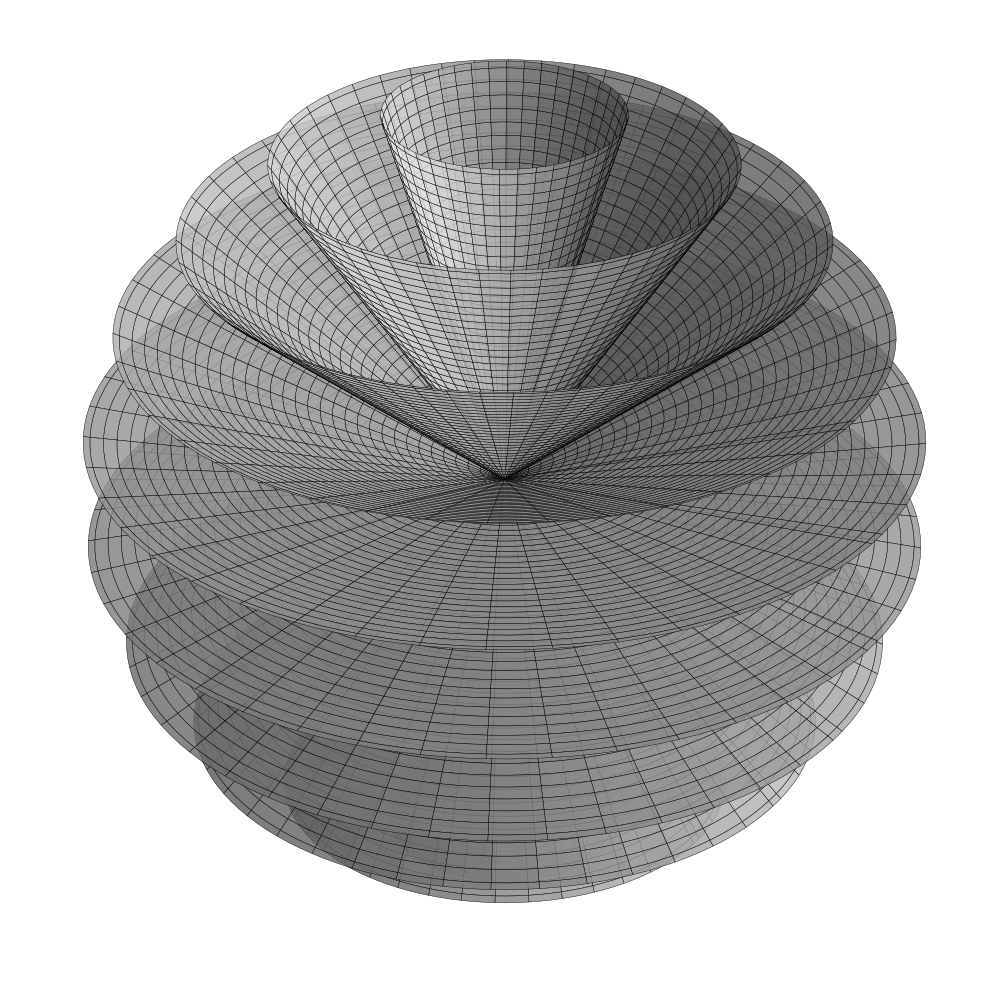}
        \caption*{(a) Conic Surfaces}
    \end{minipage}
    \begin{minipage}{0.25\textwidth}
        \centering
        \includegraphics[height=3.2cm, keepaspectratio]{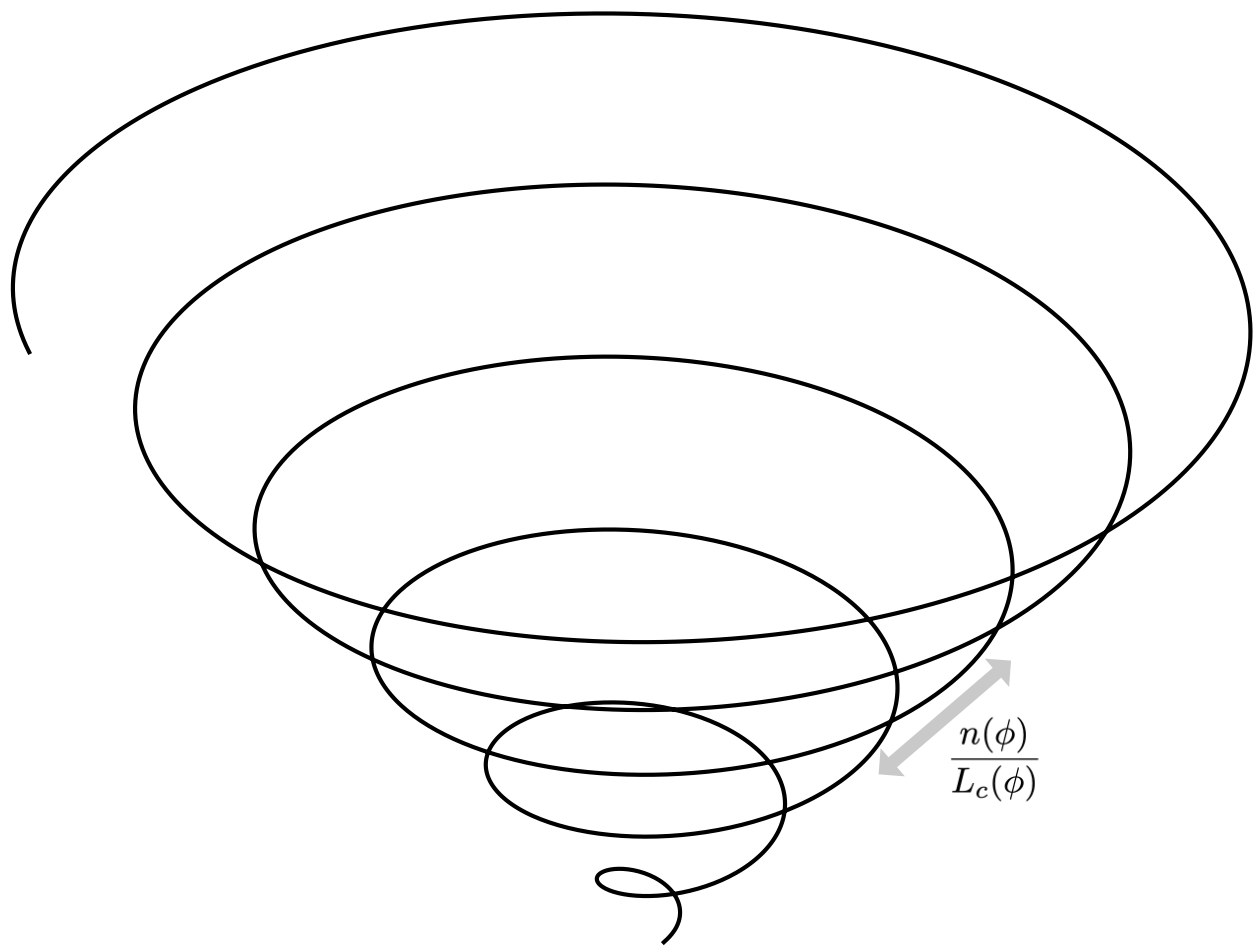}
        \caption*{(b) Conic Interleaf}
    \end{minipage}
    \caption[Nested conic surfaces.]{(a) In 3D Cones trajectories, k-space is segmented into a set of nested conic surfaces with each surface being covered by conic interleaves. (b) An interleaved conic curve at polar angle $\phi$ in k-space. The spacing between points with the same azimuthal angle is determined by the inverse of the FOV multiplied by the number of interleaves, $n(\phi)$.}
    \label{fig:conic_surfaces}
\end{figure}

One drawback of the 3D Radial trajectory is that it requires a large number of spokes to satisfy the Nyquist criterion at the periphery of k-space \cite{bernstein2004handbook, irarrazabal_fast_1995}. In 3D Cones trajectories, k-space is segmented into a set of nested conic surfaces, as illustrated in Figure \ref{fig:conic_surfaces}, with each surface being covered by conic interleaves. Since 3D Cones trajectories are comprised of these conic interleaves with twists, rather than straight radial spokes, they cover 3D k-space more efficiently than 3D Radial trajectories \cite{irarrazabal_fast_1995, larson_anisotropic_2008, gurney_design_2006}.


\subsection*{Cones Coordinates}\label{subsec:traj-cones-coord}

By only slightly abusing the notation, we represent an interleaved conic curve on a conic surface at a polar angle $\phi$:
\begin{align}
    \begin{bmatrix}
        k_x(k) \\
        k_y(k) \\
        k_z(k)
    \end{bmatrix} &= 
    \begin{bmatrix}
        k \cos \theta(k ; \phi) \sin \phi \\
        k \sin \theta(k ; \phi) \sin \phi \\
        k \cos \phi
    \end{bmatrix}, \\
    \theta(k ; \phi) &= \frac{2 \pi L_c(\phi)}{n(\phi)} k,
\end{align}
where $k$ denotes the radial distance. The term $n(\phi)$ represents the number of interleaves necessary to fill the conic surface, and $L_c(\phi)$ denotes the FOV on the conic surface:
\begin{equation}
    L_c(\phi) = \sqrt{(L_z \cos \phi)^2 + (L_r \sin \phi)^2}.\label{eq:fov_on_cone}
\end{equation}


Suppose we have an array of conic interleaves, $\{\vec{k}^{(\phi)}(t)\}$, with each residing on a distinct conic surface defined by the polar angle $\phi$. Any point within a certain radius can be uniquely represented with the set of conic templates by following these steps:
\begin{enumerate}
    \item Consider a point in a Spherical coordinate: $(\rho, \theta, \phi)$.
    \item From the set of templates, choose a template $\vec{k}^{(\phi)}(t)$ on the conic surface embracing this point.
    \item Determine the $t$ value such that $k^{(\phi)}(t) = \rho$.
    \item Rotate the template about the $k_z$-axis by $\theta - \angle \vec{k}^{(\phi)}(t)$.
\end{enumerate}
Here, $\angle \vec{k}^{(\phi)}(t)=\angle \left(k^{(\phi)}_x(t) + j k^{(\phi)}_y(t)\right)$.
\vspace*{5pt}

We introduce a new coordinate system, Cones coordinate, which uniquely determines the location of a point using the array of conic templates, $\{\vec{k}^{(\phi)}(t)\}$, and three coordinates, $(t, \theta_p, \phi_p)$. The relationship between Cones and Spherical coordinates is given as:
\begin{align}\label{eq:traj-cones-coord}
    k^{(\phi)}(t) &= \rho, \nonumber\\
    \theta_p &= \theta - \angle \vec{k}^{(\phi)}(t), \\
    \phi_p &= \phi.\nonumber
\end{align}

The Cones coordinate can be viewed as a 3D Cones trajectory that consists of infinitely many conic interleaves on continuous conic surfaces. In our approach, we design the 3D Cones trajectory by discretizing the Cones coordinate using optimized spacings to retain only the essential interleaves \cite{jang2017redesigned}. Similar to the 3D Radial trajectory, this discretization is achieved by formulating a spiral path on an ellipsoid and conducting uniform sampling along the path \cite{jang2017redesigned}.

Since the 3D Cones trajectory is primarily used for rapid volumetric imaging \cite{addy_high-resolution_2015, wu2013free}, aggressive undersampling is often desired. One strategy is to employ variable-desity sampling \cite{spielman1995magnetic, glover1999simple, lee2003fast, kim2003simple}, which allocates more samples near the k-space center to mitigate aliasing artifacts. In our approach, variable-density sampling is readily implemented by designing the conic templates accordingly. We utilize this equation to modulate the FOV with a variable-density parameter $\alpha$:
\begin{equation}
    L(k ; \phi, \alpha) = L(\phi) \left|\frac{k}{K(\phi)} \right|^{\frac{1}{\alpha} - 1}\label{eq:vd_along_radial}.
\end{equation}
Given this model, the interleaved conic curve with variable-density sampling is expressed as:
\begin{align}
    \begin{bmatrix}
        k_x(k) \\
        k_y(k) \\
        k_z(k)
    \end{bmatrix} &= 
    \begin{bmatrix}
        k \cos \theta(k ; \phi, \alpha) \sin \phi \\
        k \sin \theta(k ; \phi, \alpha) \sin \phi \\
        k \cos \phi
    \end{bmatrix}, \nonumber\\
    \theta(k ; \phi) &= \frac{2 \pi L(\phi)}{n(\phi)} \int_0^k L(\tau; \phi, \alpha) d\tau.
\end{align}

\subsection*{Discretization of the Cones coordinate}\label{subsec:traj-cones-discretization}

Consider a parametric spiral path on an ellipsoid\footnote{This ellipsoid is not necessarily in k-space or image-space, as the primary objective of this procedure is to acquire rotation angles rather than to directly construct trajectories in k-space. Here, we slightly abuse the notation of $x$, $y$, and $z$ to represent the 3D coordinate in this conceptual space.}:
\begin{align}
    \begin{bmatrix}
        p_x(u) \\
        p_y(u) \\
        p_z(u)
    \end{bmatrix}
    &=
    \begin{bmatrix}
        k_r \cos\theta_p(u) \sin\phi_p(u) \\
        k_r \sin\theta_p(u) \sin\phi_p(u) \\
        k_z \cos\phi_p(u)
    \end{bmatrix},
    \label{eq:cones_spiral_path}
\end{align}
where $u \in (0, 1)$. While $u$ is not directly related to any physical quantity, it is interpreted as time on this path in the following derivation. The time derivatives of $\theta_p(u)$ and $\phi_p(u)$ modulate the speeds of the spiral path in the azimuthal and polar directions, respectively. Our goal is to determine $\theta_p(u)$ and $\phi_p(u)$ from given imaging parameters. Subsequently, we apply the uniform sampling over $u$ at intervals of $\Delta u$ to obtain the discretized $\{\theta_p[u]\}$ and $\{\phi_p[u]\}$, where $\Delta u$ is one over the total number of readouts:
\begin{equation}
    \Delta u = \frac{1}{N}.
\end{equation} 

The duration for the spiral path to make one turn along the azimuthal direction is:
\begin{equation}
    \frac{2 \pi}{\dot{\theta_p}(u)}.\label{eq:cones_one_turn_dur}
\end{equation}

The spacing between two adjacent latitude circles can be approximated by an arc length on the ellipsoid. This spacing should be identical to the reciprocal of the FOV in the orthogonal direction:
\begin{equation}
    \underbrace{
        \frac{2 \pi}{\dot{\theta_p}(u)}
    }_{\text{duration}}
    \underbrace{
        \vphantom{\frac{2 \pi}{\dot{\theta_p}(u)}}
        \dot{\phi}_p(u)
    }_{\text{speed}}
    \underbrace{
        \vphantom{\frac{2 \pi}{\dot{\theta_p}(u)}}
        K(\phi_p(u))
    }_{\text{radius}}
    = \frac{1}{L_{90 ^{\circ}}(\phi_p(u))}.
    \label{eq:cones_dist_along_polar}
\end{equation}

In contrast to the 3D Radial trajectory, the spacing between two adjacent samples along the spiral path is not directly determined by the FOV. Instead, we need another function, $n(\phi)$, which represents the required number of interleaves on a conic surface at a polar angle $\phi$ to fully sample the surface. The number of samples during the time interval in \eqref{eq:cones_one_turn_dur} needs to match $n(\phi)$:
\begin{equation}
    \underbrace{
        \left(\frac{2 \pi}{\dot{\theta_p}(u)}\right)
    }_{\text{duration}}
    \underbrace{
        \vphantom{\left(\frac{2 \pi}{\dot{\theta_p}(u)}\right)}
        (\Delta u)
    }_{\text{interval}}
    \vphantom{(\Delta u)}^{-1} = n(\phi_p(u)).
    \label{eq:cones_nintv_along_azimuthal}
\end{equation}

From \eqref{eq:cones_dist_along_polar} and \eqref{eq:cones_nintv_along_azimuthal}, we derive the differential equations for $\phi_p(u)$ and $\theta_p(u)$:
\begin{align}
    \dot{\phi}_p(u) &= \frac{N}{K(\phi_p(u)) n(\phi_p(u)) L_{90^{\circ}}(\phi_p(u))},
    \label{eq:cones_diffeq_phi}\\
    \dot{\theta}_p(u) &= \frac{2 \pi N}{n(\phi_p(u))},\label{eq:cones_diffeq_theta}
\end{align}
\noindent where $N = (\Delta u)^{-1}$ represents the number of readouts.

Because \eqref{eq:cones_diffeq_phi} has the form of \eqref{eq:diffeq_problem}, we apply Algorithm \ref{alg:diffeq} to compute $\phi_p(u)$. Like the 3D Radial trajectory, the elliptical $K$ and FOV models, described in Table~\ref{tab:notation}, are employed for anisotropic FOV and/or resolution. Algorithm \ref{alg:cones_full} outlines how to formulate the spiral path on the ellipsoid.

A discretized point in the Cones coordinate, $(\theta_i, \phi_i)$, corresponds to the $i$-th conic interleaf in k-space, $\vec{k}^{(i)}(t)$, synthesized by rotating a template, $\vec{k}^{(\phi_i)}(t)$, around the $k_z$-axis by $\theta_i$.

\begin{algorithm}[t!]
    \centering
    \caption{Spiral Path on Ellipsoid for 3D Cones Trajectories}
    \begin{tabular}{p{0.05\textwidth}p{0.38\textwidth}}
    \toprule
    \textbf{Input} & Models for the FOV and $K$, i.e., $L(\phi)$ and $K(\phi)$. \\
    \textbf{Output} & $\phi_p(u)$ and $\theta_p(u)$. \\
    \midrule
    \textbf{Step 1} & Construct templates on multiple conic surfaces and estimate the number of interleaves $n(\phi)$. \\
    \textbf{Step 2} & Calculate the number of readouts $N$: 
    \[
    N = \int_0^{\pi} K(\phi) n(\phi) L_{90^{\circ}}(\phi) d\phi.
    \] \\
    \textbf{Step 3} & Determine $u(\phi)$ using the cumulative summation:
    \[
    u(\phi) = \frac{1}{N}\int_0^\phi K(\tau) n(\tau) L_{90^{\circ}}(\tau) d\tau.
    \] \\
    \textbf{Step 4} & Compute the inverse of $u(\phi)$ over $u \in (0, 1)$ to get $\phi_p(u)$. \\
    \textbf{Step 5} & Determine $\theta_p(u)$ using the cumulative summation:
    \[
    \dot{\theta}_p(u) = \frac{2 \pi N}{n(\phi_p(u))}.
    \] \\[-8pt]
    \bottomrule
    \end{tabular}
    \label{alg:cones_full}
\end{algorithm}

\subsection*{Density Compensation Factor}\label{subsec:traj-cones-dcf}

As with the 3D Radial trajectory, the DCF is proportional to both the Jacobian determinant associated with the transformation of the Cones coordinate to the Cartesian coordinate \cite{hoge_density_1997} and the inverse of the distribution of discretized samples.

Because the relationship between the Cones and Spherical coordinates is established in \eqref{eq:traj-cones-coord}, and because the transformation from Spherical coordinate to Cartesian coordinate is well-known, we compute the Jacobian determinant by using the Spherical coordinate as an intermediary.

Specifically, the Jacobian determinant between the Cones coordinate and Spherical coordinate is:
\begin{align}
    \det(J_{\text{Cones-to-Sph}}) &= 
    \begin{vmatrix}  
        \dfrac{\partial \rho}{\partial t} & \dfrac{\partial \theta}{\partial t} & \dfrac{\partial \phi}{\partial t}\nonumber\\[10pt]
        \dfrac{\partial \rho}{\partial \theta_p} & \dfrac{\partial \theta}{\partial \theta_p} & \dfrac{\partial \phi}{\partial \theta_p}\nonumber\\[10pt]
        \dfrac{\partial \rho}{\partial \phi_p} & \dfrac{\partial \theta}{\partial \phi_p} & \dfrac{\partial \phi}{\partial \phi_p}
    \end{vmatrix}\nonumber\\[10pt]
    &= \frac{\partial \rho}{\partial t} = \frac{\gamma}{2 \pi} \frac{(\vec{g} \cdot \vec{k})[t]}{k[t]}.
\end{align}
The Jacobian determinant between the Spherical coordinate and the Cartesian coordinate is:
\begin{equation}
    \det(J_{\text{Sph-to-Cart}}) = \rho^2 \sin \phi = k^2[t] \sin \phi_p[u].
\end{equation}

\noindent Thus, the Jacobian determinant associated with the transformation of the Cones coordinate to the Cartesian coordinate is:
\begin{align}
    &|\det(J_{\text{Cones-to-Cart}})| \nonumber \\
    &= |\det(J_{\text{Cones-to-Sph}})|
    \cdot |\det(J_{\text{Sph-to-Cart}})| \\
    &= \frac{\gamma}{2 \pi} k[t]|(\vec{g}\cdot\vec{k})[t]| \sin \phi_p[u].\nonumber 
\end{align}

The denominator of $\dot{\phi}_p(u)$ in \eqref{eq:cones_diffeq_phi} corresponds to the probability density function of discretized samples, as shown in \eqref{eq:p_and_g}. Hence, the final DCF is:
\begin{align}
    d[t, u] &\propto \frac{|\det(J_{\text{Cones-to-Cart}})|}{g[u]} \nonumber\\
    &= \frac{|\det(J_{\text{Cones-to-Cart}})|}{K(\phi_p(u)) n(\phi_p(u)) L_{90^{\circ}}(\phi_p(u))}\label{eq:cones-dcf}\\
    &= \frac{\gamma}{2 \pi} \frac{k[t]|(\vec{g}\cdot\vec{k})[t]| \sin \phi_p[u]}{K(\phi_p[u]) n(\phi_p[u]) L_{90^{\circ}}(\phi_p[u])}.\nonumber
\end{align}

%% file: sections/spiral.tex
\section*{3D Stack-of-spiral trajectories}\label{sec:spiral}

Stack-of-spiral trajectories consist of multiple spiral interleaves on planar surfaces parallel to the $k_x$-$k_y$ plane \cite{irarrazabal_fast_1995,thedens_fast_1999}. In contrast to the 3D Radial and Cones coordinates that are similar to Spherical coordinates, the analytic coordinate of stack-of-spirals trajectories is closely related to Cylindrical coordinates. Nevertheless, the structure of the derivation and the resulting algorithm are similar to those for the 3D Radial and 3D Cones trajectories. 

\subsection*{Stack-of-Spirals Coordinates}\label{subsec:traj-spiral-coord}

Consider a set of spiral interleaves, $\{\vec{k}^{(z)}(t)\}$, each associated with a distinct horizontal surface at the $k_z$-coordinate (denoted here as $z$)\footnote{For simplicity, we slightly abuse the notation of $z$ in this section.}. Any point within a certain elevation range in k-space can be uniquely represented using this procedure:
\begin{enumerate}
    \item Consider this point in the Cylindrical coordinate: $(r, \theta, z)$.
    \item Select a template $\vec{k}^{(z)}(t)$ that resides on the horizontal surface at $z$ from the set of templates.
    \item Determine the $t$ value for which $k^{(z)}_r(t) = r$.
    \item Rotate the selected template about the $k_z$-axis by an angle of $\theta - \angle \vec{k}^{(z)}(t)$.
\end{enumerate}

Analogous to the 3D Radial and Cones coordinates, we introduce a new coordinate system, Stack-of-Spirals coordinate, which uniquely determines the location of a point using an array of spiral templates, $\{\vec{k}^{(z)}(t)\}$, and three coordinates: $(t, \theta_p, z_p)$. These coordinates relate to the Cylindrical coordinate as follows:
\begin{align}
    k^{(z)}_r(t) &= r, \nonumber\\
    \theta_p &= \theta - \angle \vec{k}^{(z)}(t), \\
    z_p &= z.\nonumber
\end{align}

This new coordinate can be viewed as a Stack-of-Spirals trajectory comprising infinitely many spiral interleaves across continuous horizontal surfaces. 
We design the Stack-of-Spirals trajectory by discretizing the Stack-of-Spirals coordinate using optimized spacings to retain only the essential spiral interleaves for given imaging parameters. Similar to the 3D Radial and 3D Cones trajectories, this discretization is achieved by formulating a spiral path and conducting uniform sampling. However, in this case, we need to consider a cylindrical surface rather than an ellipsoidal surface.

To enhance sampling efficiency, we incorporate spherical designs for stack-of-spiral trajectories \cite{thedens_fast_1999, asslander2013single} by creating templates with varied radial extents as a function of z position. Algorithm \ref{alg:stack_spherical} summarizes how to determine a radial resolution on a horizontal surface. 

\begin{algorithm}[t!]
    \centering
    \caption{Radial Resolution for Spherical Stack-of-Spirals Trajectories}
    \begin{tabular}{p{0.05\textwidth}p{0.38\textwidth}}
    \toprule
    \textbf{Input} & Minimum resolution $\Delta_{\min}$ and spherical $K(\phi)$ model. \\
    \textbf{Output} & Radial resolution $\Delta_r$ for given $k_z$. \\
    \midrule
    \textbf{Step 1} & Compute the polar angle $\phi$: 
    \[
    \phi = \cos^{-1} \left( \frac{K_z k_{z}}{\sqrt{(K_z K_r)^2 + (K_z k_{z})^2 - (K_r k_{z})^2}} \right).
    \] \\
    \textbf{Step 2} & Calculate the maximum radius in k-space:
    \[
    k_{r, \max}(\phi) = | K(\phi) \sin(\phi) |
    \]
    \[
    k_{r, \max}(\phi) = \frac{K_r K_z | \sin(\phi) |}{\sqrt{( K_z \sin\phi)^2 + ( K_r \cos\phi)^2}}.
    \] \\
    \textbf{Step 3} & Determine the radial resolution:
    \[
    \Delta_r(\phi) = \min \left(\frac{1}{2 k_{r, \max}(\phi)}, \Delta_{\min} \right).
    \] \\[-8pt]
    \bottomrule
    \end{tabular}
    \label{alg:stack_spherical}
\end{algorithm}

\subsection*{Discretization of the Stack-of-Spirals Coordinate}\label{subsec:traj-spiral-discretization}

Imagine a parametric spiral path on a cylinder\footnote{This cylinder is not necessarily in k-space or image-space, as the primary objective of this procedure is to acquire rotation angles rather than to directly construct trajectories in k-space. Here, we slightly abuse the notation of $x$, $y$, and $z$ to represent the 3D coordinate in this conceptual space.} with a unit radius:
\begin{align}
    \begin{bmatrix}
        p_x(u) \\
        p_y(u) \\
        p_z(u)
    \end{bmatrix}
    &=
    \begin{bmatrix}
        \cos\theta_p(u) \\
        \sin\theta_p(u) \\
        z_p(u)
    \end{bmatrix},
    \label{eq:stack_spiral_path}
\end{align}
where $u \in (0, 1)$. While $u$ is not directly related to any physical quantity, it is interpreted as time on this path in the following derivation. The time derivatives of $\theta_p(u)$ and $z_p(u)$ dictate the velocities in the azimuthal and z-directions, respectively. We determine $\theta_p(u)$ and $z_p(u)$ based on prescribed parameters and subsequently employ uniform sampling with intervals of $\Delta u$ over $u$ to acquire the discretized $\{\theta_p[u]\}$ and $\{z_p[u]\}$.

The duration required for the spiral path to complete one azimuthal rotation is:
\begin{equation}
    \frac{2 \pi}{\dot{\theta_p}(u)}. \label{eq:stack_duration}
\end{equation}

The spiral's progression along the z-axis during this time period is set as the inverse of the FOV in the z-direction:
\begin{equation}
    \underbrace{
        \frac{2 \pi}{\dot{\theta_p}(u)}
    }_{\text{duration}}
    \underbrace{
        \vphantom{\frac{2 \pi}{\dot{\theta_p}(u)}}
        \dot{z}_p(u)
    }_{\text{speed}} = \frac{1}{L_z(z_p(u))}.
    \label{eq:stack_dist_along_z}
\end{equation}

The number of samples during this time interval should match $n(z_p(u))$, where $n(z_p(u))$ denotes the number of interleaves necessary for sampling a horizontal plane at $k_z = z_p(u)$:
\begin{equation}
    \underbrace{
        \frac{2 \pi}{\dot{\theta_p}(u)}
    }_{\text{duration}}
    \underbrace{
        \vphantom{\frac{2 \pi}{\dot{\theta_p}(u)}}
        (\Delta u)
    }_{\text{interval}}
    \vphantom{(\Delta u)}^{-1} = n(z_p(u)).
    \label{eq:stack_along_ring}
\end{equation}

From \eqref{eq:stack_dist_along_z} and \eqref{eq:stack_along_ring}, we derive the differential equations for $z_p(u)$ and $\theta_p(u)$:
\begin{align}
    \dot{z}_p(u) &= \frac{N}{n(z_p(u)) L_z(z_p(u))},\label{eq:stack_diffeq_z}\\
    \dot{\theta}_p(u) &= \frac{2 \pi N}{n(z_p(u))}.\label{eq:stack_diffeq_theta}
\end{align}
Here, $N = (\Delta u)^{-1}$ represents the total number of readouts. Given that \eqref{eq:stack_diffeq_z} has the form of \eqref{eq:diffeq_problem}, we utilize Algorithm \ref{alg:diffeq} to determine $z_p(u)$. The procedure for computing $z_p(u)$ and $\theta_p(u)$ for the Stack-of-Spirals trajectory is summarized in Algorithm \ref{alg:stack_full}.

A discretized point in the Stack-of-Spirals coordinate, $(\theta_i, z_i)$, corresponds to the $i$-th spiral interleaf in k-space, $\vec{k}^{(i)}(t)$, synthesized by rotating a template, $\vec{k}^{(z_i)}(t)$, around the $k_z$-axis by $\theta_i$.

\begin{figure}[!htb]
    \centering
    \begin{minipage}{0.22\textwidth}
        \centering
        \includegraphics[width=\linewidth]{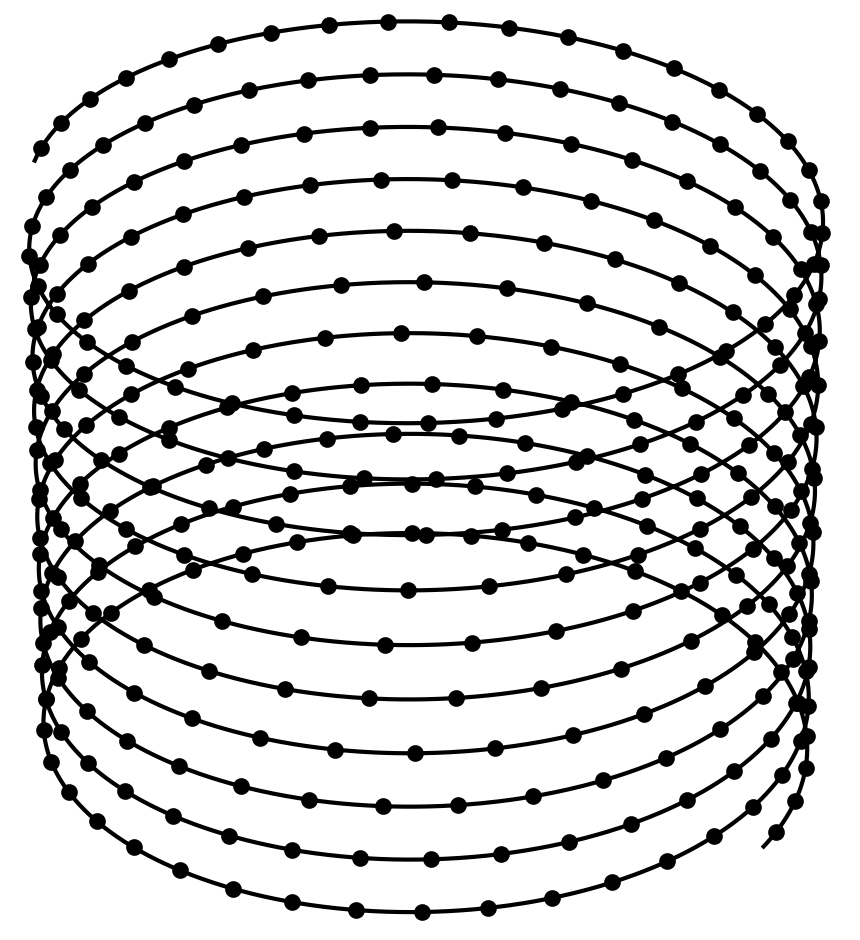}
        \caption*{(a) Default}
    \end{minipage}
    \begin{minipage}{0.22\textwidth}
        \centering
        \includegraphics[width=\linewidth]{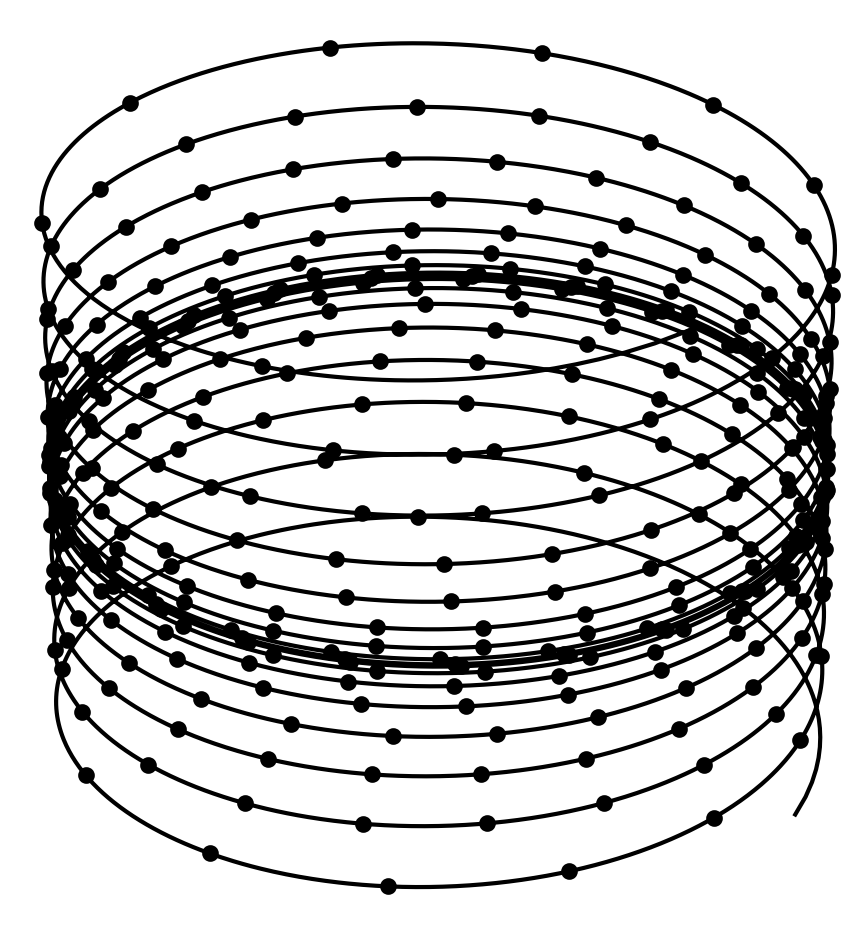}
        \caption*{(b) Spherical and variable-density}
    \end{minipage}
    \caption[Spiral paths on a cylinder.]{Spiral paths on a cylinder. (a) without the spherical and variable-density design, (b) with the spherical design and variable-density ($\alpha_z=2.5$) along z-direction. Dots represent discretized points. In (b), the intervals between points at both ends of the spiral path are larger due to the spherical design. Additionally, spacings along the z-direction are denser near the equator, reflecting the variable-density design.}
    \label{fig:stack_spiral_path}
\end{figure}

As demonstrated in \eqref{eq:stack_diffeq_z}, $L_z(z)$ directly modulates the speed of the spiral path along the z-axis. By adjusting $L_z(z)$, the spiral path can progress more slowly near the equator, resulting in a denser concentration of samples around $k_z=0$. Analogous to \eqref{eq:vd_along_radial}, we determine the FOVs along the radial direction and the z-axis for variable-density sampling as follows:
\begin{align}
    L_r(k ; \phi, \alpha_r) &= L_r \left|\frac{k}{2 \Delta_r(\phi)} \right|^{\frac{1}{\alpha_r} - 1}\label{eq:vd_along_r},\\
    L_z(z ; \alpha_z) &= L_z(z) \left|\frac{z}{k_{\max, z}} \right|^{\frac{1}{\alpha_z} - 1}\label{eq:vd_along_z}.
\end{align}
The spiral path with the variable-density parameter $\alpha_z=2.5$ is depicted in Figure \ref{fig:stack_spiral_path}.

\begin{algorithm}[t!]
    \centering
    \caption{Spiral Path on Cylinder for Stack-of-Spirals Trajectories}
    \begin{tabular}{p{0.05\textwidth}p{0.38\textwidth}}
    \toprule
    \textbf{Input} & FOV and $k_{\max}$ with respect to $k_z$, i.e., $L_z(z)$ and $k_{\max}(z)$. \\
    \textbf{Output} & $z_p(u)$ and $\theta_p(u)$. \\
    \midrule
    \textbf{Step 1} & Construct multiple templates for z-planes and estimate the number of interleaves $n(z)$. \\
    \textbf{Step 2} & Calculate the number of readouts $N$:
    \[
    N = \int_{-k_{\max, z}}^{+k_{\max, z}} n(z) L_z(z) ~ dz.
    \] \\
    \textbf{Step 3} & Determine $u(z)$ using the cumulative summation:
    \[
    u(z) = \frac{1}{N} \int_{-k_{\max, z}}^{z} n(\tau) L_z(\tau) ~ d\tau.
    \] \\
    \textbf{Step 4} & Compute the inverse of $u(z)$ over $u \in (0, 1)$ to get $z_p(u)$. \\
    \textbf{Step 5} & Determine $\theta_p(u)$ using the cumulative summation:
    \[
    \dot{\theta}_p(u) = \frac{2 \pi N}{n(z_p(u))}.
    \] \\[-8pt]
    \bottomrule
    \end{tabular}
    \label{alg:stack_full}
\end{algorithm}

\subsection*{Density Compensation Factor}\label{subsec:traj-spiral-dcf}

Similar to 3D Radial and 3D Cones trajectories, an analytical DCF is derived using the Jacobian determinant associated with the transformation of the Stack-of-Spirals coordinate to the Cartesian coordinate \cite{hoge_density_1997}, in combination with the probability density function of discretized samples.

First, we compute the Jacobian determinant between the Stack-of-Spirals coordinate and the Cylindrical coordinate:
\begin{align}
    \det(J_{\text{Sp-to-Cyl}}) &= 
    \begin{vmatrix}  
        \dfrac{\partial r}{\partial t} & \dfrac{\partial \theta}{\partial t} & \dfrac{\partial z}{\partial t}\\[10pt]
        \dfrac{\partial r}{\partial \theta_p} & \dfrac{\partial \theta}{\partial \theta_p} & \dfrac{\partial z}{\partial \theta_p}\\[10pt]
        \dfrac{\partial r}{\partial z_p} & \dfrac{\partial \theta}{\partial z_p} & \dfrac{\partial z}{\partial z_p}
    \end{vmatrix}\nonumber\\[10pt]
    &= \frac{\partial r}{\partial t} = \frac{\gamma}{2 \pi} \frac{(\vec{g}_r \cdot \vec{k}_r)[t]}{k_r[t]}.
\end{align}

The Jacobian determinant between the Cylindrical coordinate and the Cartesian coordinate is equivalent to the radial distance:
\begin{equation}
    \det(J_{\text{Cyl-to-Cart}}) = r = k_r[t].
\end{equation}

Consequently, the desired Jacobian determinant is:
\begin{align}
    \det(J_{\text{Sp-to-Cart}}) &= \det(J_{\text{Sp-to-Cyl}}) \cdot \det(J_{\text{Cyl-to-Cart}})\nonumber\\
    &= \frac{\gamma}{2 \pi}(\vec{g}_r \cdot \vec{k}_r)[t].
\end{align}

The probability density function of discretized samples is proportional to the denominator of $\dot{z}_p(u)$, as shown in \eqref{eq:p_and_g}. The final DCF expression is, therefore, derived as:
\begin{align}
    d[t, u] &\propto \frac{|\det(J_{\text{Sp-to-Cart}})|}{g[n]}\nonumber\\
    &= \frac{|\det(J_{\text{Sp-to-Cart}})|}{n(z_p(u)) L_z(z_p(u))}\label{eq:spiral-dcf}\\
    &= \frac{\gamma}{2 \pi} \frac{|(\vec{g}_r \cdot \vec{k}_r)[t]|}{n(z_p[u]) L_{z}(z_p[u])}.\nonumber
\end{align}

%% file: sections/result.tex
\section*{Phantom and In-Vivo Studies}

Data were acquired on a 1.5T GE Signa scanner using a GE 8-channel cardiac coil. For the phantom study, an American College of Radiology (ACR) phantom was used. A split Bregman algorithm with total variation regularization \cite{goldstein_split_2009} was applied for reconstructions.

\subsection*{3D Cones Trajectories}

\begin{figure}[!htb]
    \centering
    \begin{minipage}{0.23\textwidth}
        \centering
        \includegraphics[width=\linewidth]{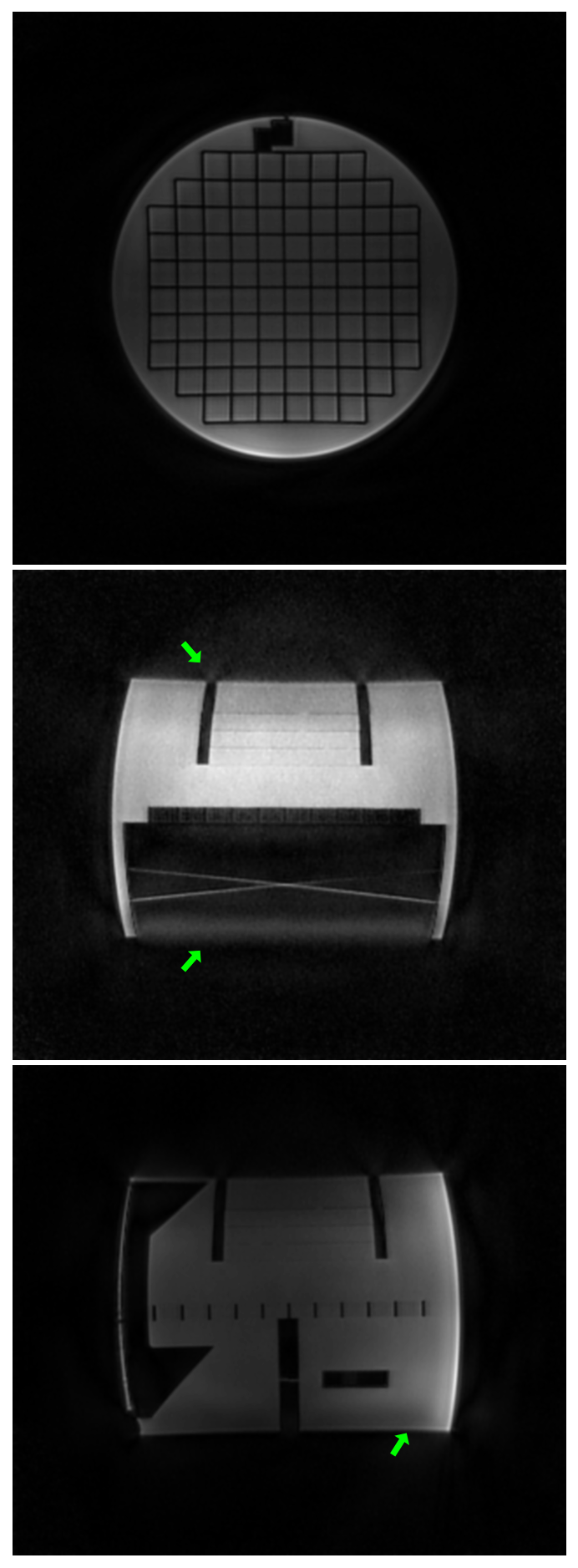}
        \caption*{(a) conventional}
    \end{minipage}
    \begin{minipage}{0.23\textwidth}
        \centering
        \includegraphics[width=\linewidth]{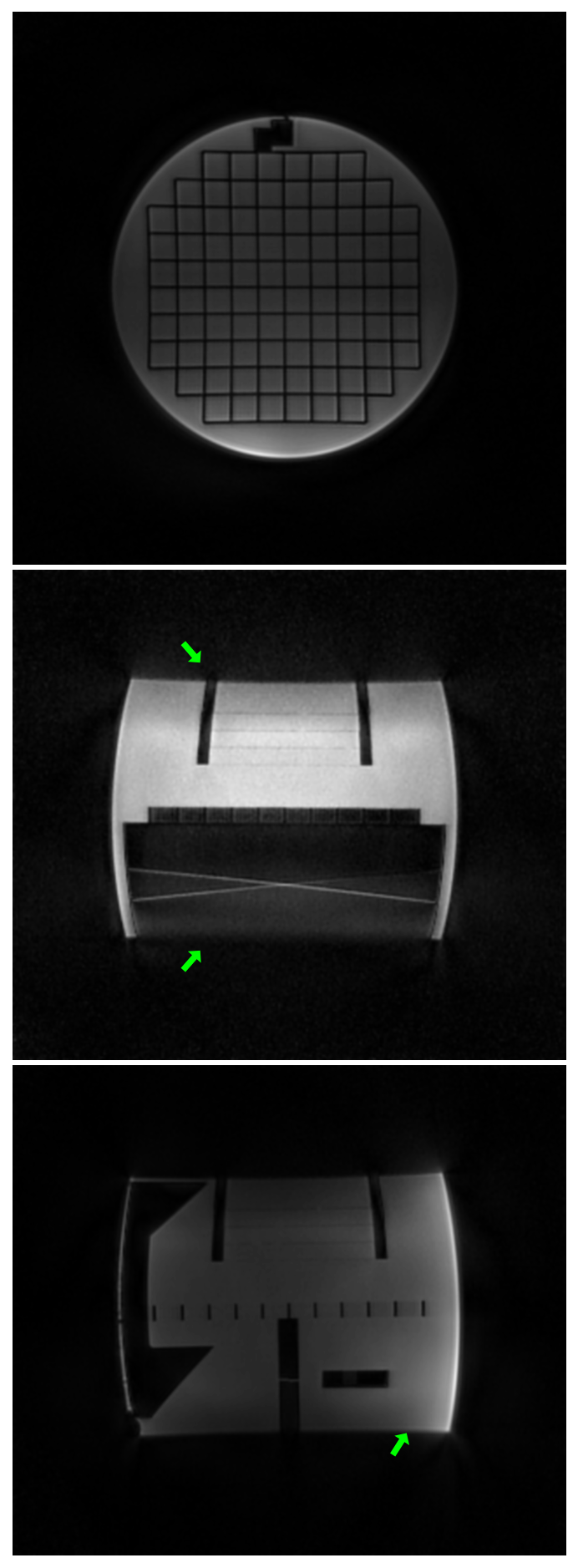}
        \caption*{(b) proposed}
    \end{minipage}
    \caption[Phantom Study (high-resolution, fully sampled case).]{ACR phantom study using fully sampled 3D Cones trajectories. (a) Conventional and (b) proposed approach. The proposed trajectory achieved more accurate reconstruction in the outer regions, while both methods yielded similar results near the center.}
    \label{fig:cones-phantom-highres}
\end{figure}

We designed fully sampled 3D Cones trajectories using the conventional \cite{gurney_design_2006} and the proposed method with the following imaging parameters: a readout duration of 2.8 ms, an isotropic resolution of 1.2 mm, an anisotropic FOV of (28, 28, 14) cm, a maximum gradient amplitude of 39 mT/m, and a maximum slew rate of 145 mT/m/ms. The proposed approach demonstrated a slight improvement in efficiency, requiring 8,862 readouts compared to 9,210 readouts for the conventional method, corresponding to a reduction of approximately 3.78\%. To match the number of readouts, a small variable-density parameter ($\alpha=1.065$) was used for the proposed method in the following comparisons. As illustrated in Figure~\ref{fig:cones-phantom-highres}, the proposed trajectory provided more accurate reconstruction in the outer regions, while both trajectories produced very similar results near the center.

\begin{figure}[!htb]
    \centering
    \begin{minipage}{0.23\textwidth}
        \centering
        \includegraphics[width=\linewidth]{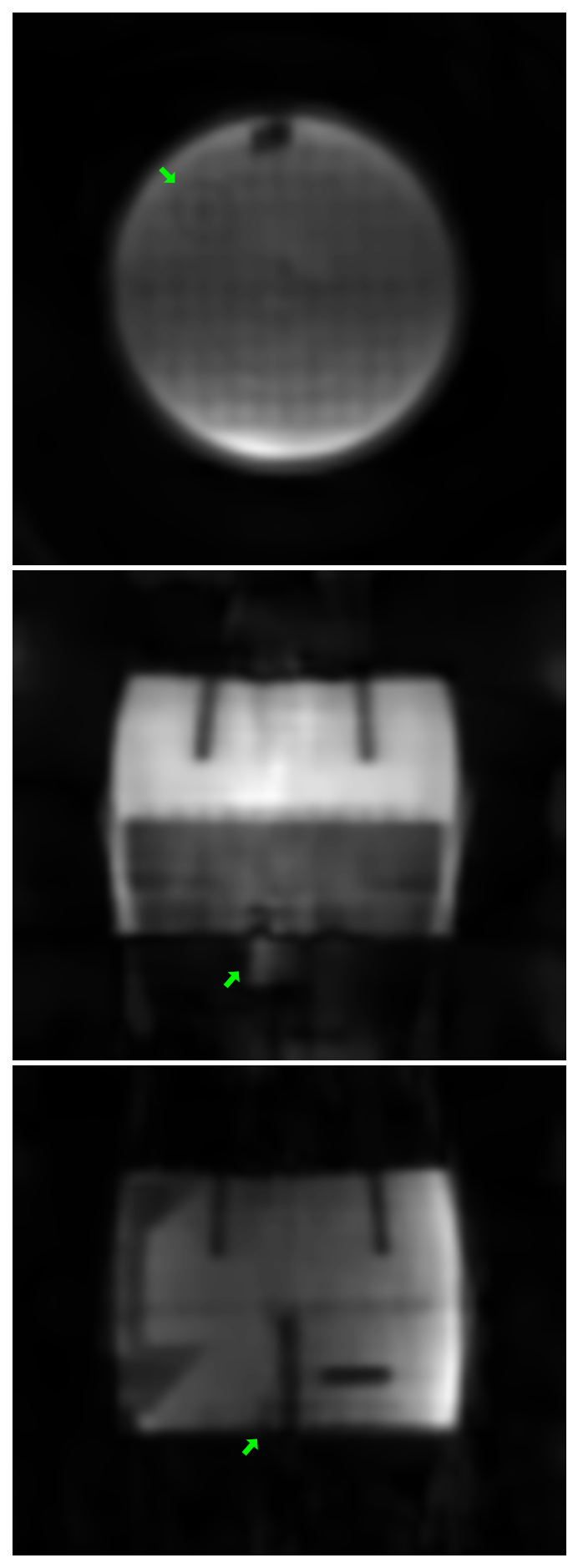}
        \caption*{(a) conventional}
    \end{minipage}
    \begin{minipage}{0.23\textwidth}
        \centering
        \includegraphics[width=\linewidth]{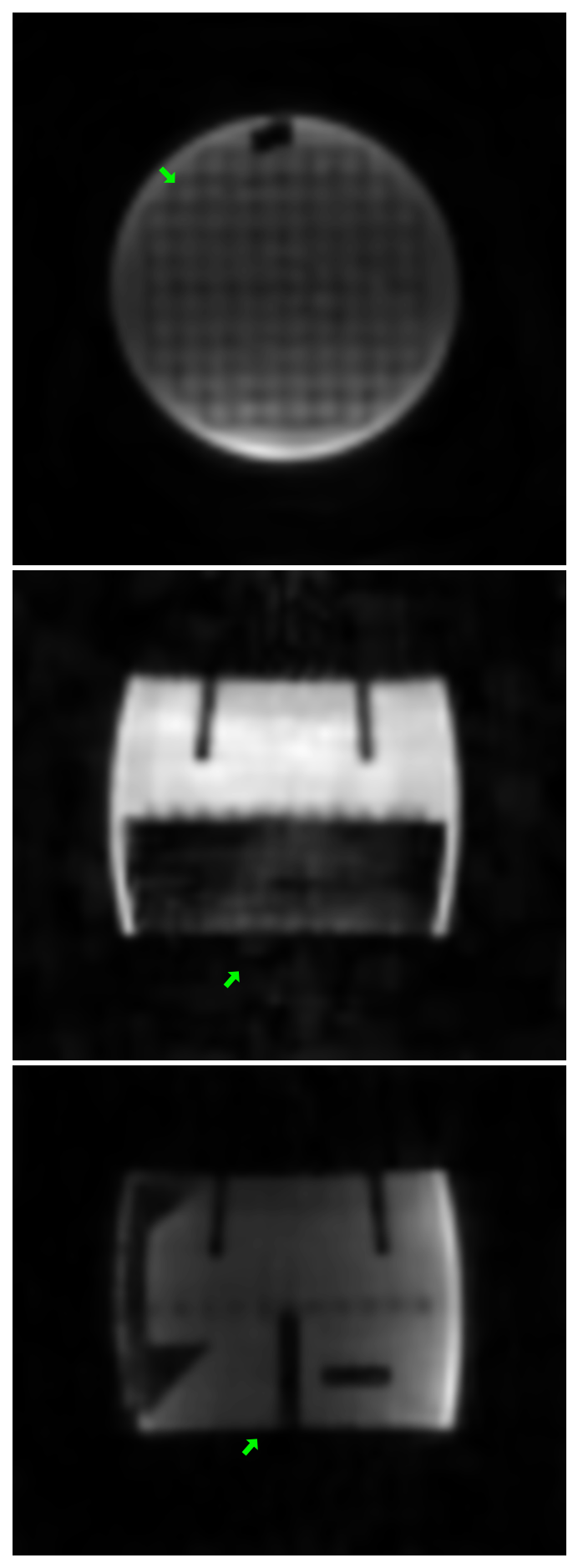}
        \caption*{(b) proposed}
    \end{minipage}
    \caption[Phantom Study (low-resolution, undersampled case).]{ACR phantom study using undersampled 3D Cones trajectories. (a) Conventional and (b) proposed approach. The proposed trajectory significantly reduced z-directional aliasing artifacts, demonstrating the effectiveness of variable-density sampling in highly undersampled scenarios.}
    \label{fig:cones-phantom-inav}
\end{figure}

We also designed a highly undersampled 3D Cones trajectory for 3D image-based navigators (3D iNAVs) using the following imaging parameters: 32 readouts, a readout duration of 2.8 ms, an isotropic resolution of 4.4 mm, an anisotropic FOV of (28, 28, 14) cm, and a variable-density parameter of 2.25. As illustrated in Figure~\ref{fig:cones-phantom-inav}, the proposed method exhibited a noticeable improvement over the conventional trajectory \cite{addy20173d}, substantially reducing z-directional aliasing artifacts. These results demonstrate that variable-density sampling, incorporated in the proposed method as shown in \eqref{eq:vd_along_radial}, is highly effective in undersampled scenarios.

\subsection*{3D Stack-of-Spiral Trajectories}

We tested the feasibility of capturing a 3D coronary artery MR image within a reasonably short breath-hold period. To achieve this goal, we designed highly undersampled 3D Spherical Stack-of-Spiral trajectories using the following imaging parameters: 360 readouts, a readout duration of 3.2 ms, an anisotropic resolution of (1.2, 1.2, 1.5) mm, an anisotropic FOV of (28, 28, 3) cm, a maximum gradient amplitude of 39 mT/m, and a maximum slew rate of 145 mT/m/ms. During each heartbeat, data were acquired along 20 spiral interleaves using a balanced steady-state free precession (bSSFP) sequence (TE/TR = 0.7/5.0 ms).

Three combinations of the variable-density parameters were tested: (a) no variable-density ($\alpha_r=1$, $\alpha_z=1$), (b) variable-density only in the radial direction ($\alpha_r=1.5$, $\alpha_z=1$), and (c) variable-density in both the radial and z-directions ($\alpha_r=1.5$, $\alpha_z=2.25$).

\begin{figure}[!htb]
    \centering
    \begin{minipage}{0.15\textwidth}
        \centering
        \includegraphics[width=\linewidth]{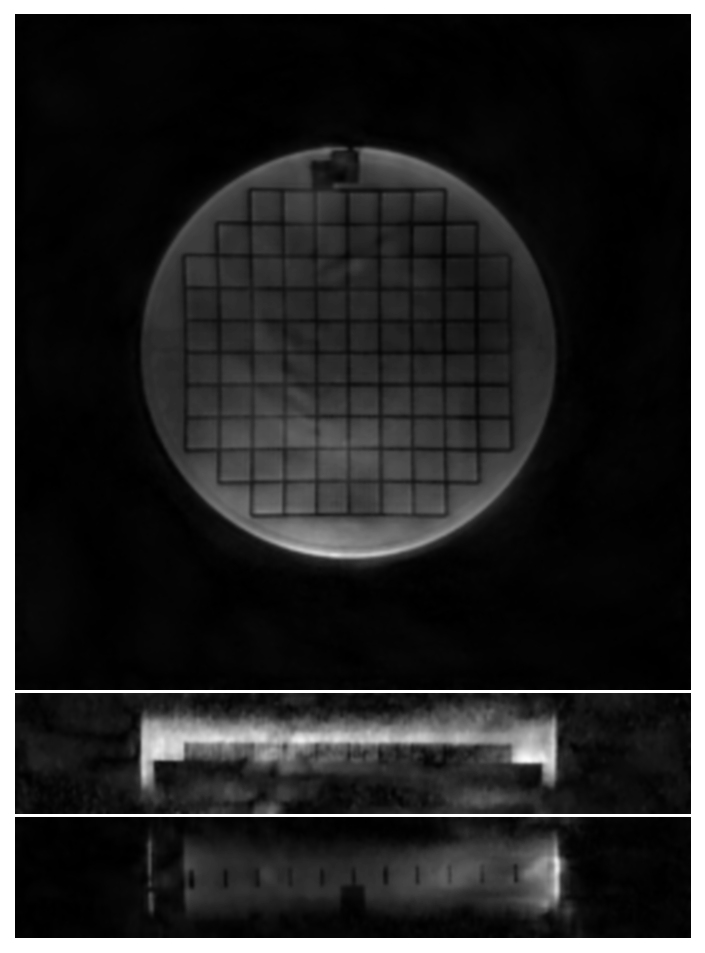}
        \caption*{(a) $\alpha_r\!=\!1$, $\alpha_z\!=\!1$}
    \end{minipage}
    \begin{minipage}{0.15\textwidth}
        \centering
        \includegraphics[width=\linewidth]{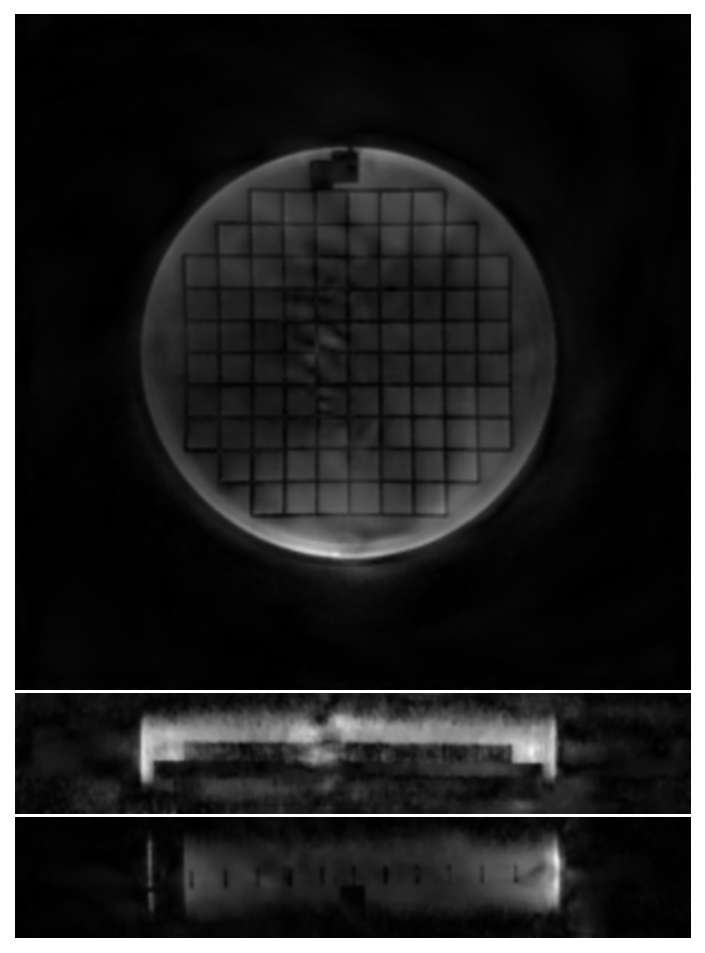}
        \caption*{(b) $\alpha_r\!=\!1.5$, $\alpha_z\!=\!1$}
    \end{minipage}
    \begin{minipage}{0.15\textwidth}
        \centering
        \includegraphics[width=\linewidth]{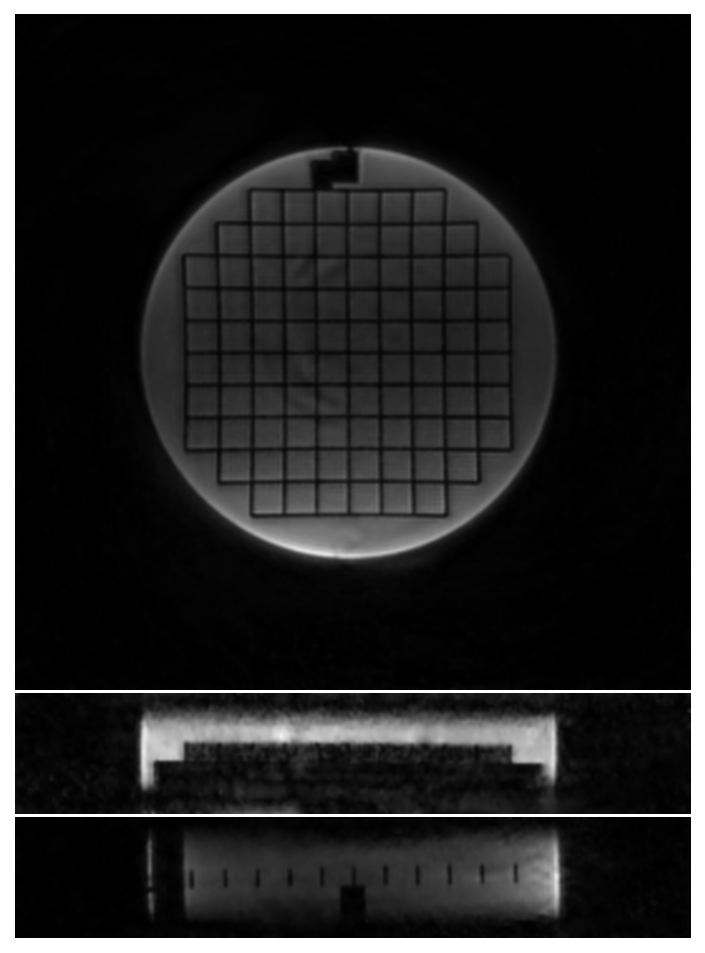}
        \caption*{(c) $\alpha_r\!=\!1.5$, $\alpha_z\!=\!2.25$}
    \end{minipage}
    \caption[Phantom Study (Stack-of-Spirals).]{ACR phantom study using undersampled Stack-of-Spirals trajectories with different variable-density settings. (a) no variable-density, (b) radial ariable-density only, and (c) radial and z-direction variable-density. Variable-density sampling in both directions substantially improved image quality, whereas using it only in the radial direction offers no notable benefit.}
    \label{fig:spiral-phantom}
\end{figure}

When 3D Stack-of-Spiral trajectories are employed in a highly undersampled scenario, z-directional aliasing artifacts are more severe than in-plane artifacts. This is because the low-frequency regions on each $k_z$-plane are repeatedly acquired along spiral interleaves, providing some robustness against undersampling in the radial direction. In contrast, in the $k_z$-direction, the low-frequency region is not sampled more frequently than the high-frequency region, making it more susceptible to z-directional aliasing artifacts. Figure~\ref{fig:spiral-phantom} illustrates the reconstructed ACR phantoms and confirms this reasoning. No improvement was observed with variable-density sampling in the radial direction only. Variable-density sampling in both directions produced substantially better results than the other trajectories.

%% file: sections/discussion.tex
\section*{Discussion}

The proposed approach derives analytic formulae to compute the number of readouts based on prescribed parameters, as given in Algorithms \ref{alg:3dpr_full}, \ref{alg:cones_full}, and \ref{alg:stack_full} for 3DPR, 3D Cones, and 3D Stack-of-spiral trajectories, respectively. This is particularly useful in practice because it is typically desirable to specify the undersampling ratio (often referred to as the acceleration factor), and the number of readouts is often constrained to limit the total scan time. In the released code, a bisectional search algorithm adjusts the FOV to match a desired number of readouts for prescribed parameters.

In conventional design methods for 3D Cones and 3D Stack-of-spiral trajectories, interleaves are allocated in 3D k-space using a two-step process: first, determining the number of discretized conic or planar surfaces, and then assigning the number of readouts to each surface. However, balancing these two quantities is often nontrivial, especially when the number of readouts is small. In the proposed approach, conic and spiral interleaves are smoothly distributed along the derived spiral path, eliminating the need to separately determine the number of surfaces and the number of readouts per surface.

%% file: sections/acknowledgement.tex
\section*{Acknowledgments}

This study was supported by NIH Grant R01 HL127039. The authors thank Dr. Bob S. Hu for his valuable discussions on 3D Stack-of-Spirals trajectories and for granting access to the 1.5T MRI scanner used in this study. Portions of this work were derived from K.E.J.'s doctoral thesis submitted to Stanford University, available at \href{https://purl.stanford.edu/sc219dv4593}{https://purl.stanford.edu/sc219dv4593}.

%% file: sections/diffeq.tex
\section*{Appendix 1}

Recall the differential equation in \eqref{eq:diffeq_problem}:
\begin{equation}
    \frac{df}{du} = \dot{f}(u) = \frac{N}{g(f(u))},
\end{equation}
where $u$ is a bounded variable ranging from 0 to 1, $g(\cdot)$ is a known function, $N$ denotes a constant, and $f(u)$ is a function with a positive, monotonically increasing inverse, $u(f)$. We also assume that $f(u)$ is bounded such that $f(u) \in (f_{\min}, f_{\max})$.

Given the constraints on $u(f) \in (0, 1)$, we interpret it as a cumulative distribution function:
\begin{align}
    u(f) &= \frac{1}{Z} \int_{f_{\min}}^{f} p(\tau) d\tau,\label{eq:t_of_f}\\
    Z &= \int_{f_{\min}}^{f_{\max}} p(\tau) d\tau, \label{eq:def_z}
\end{align}
\noindent where $p(f)$ is a probability density function and $Z$ is its normalization factor.

Differentiating both sides of \eqref{eq:t_of_f}, we represent $\dot{u}(f)$ using the probability density function:
\begin{equation}
    \frac{du}{df} = \frac{1}{Z} p(f). \label{eq:dt_over_df_1}
\end{equation}

Using the ``Reciprocal Derivative Relationship,'' we find another expression for $\dot{u}(f)$:
\begin{align}
    \frac{du}{df} &= \frac{1}{\left(\dfrac{df}{du}\right)} \\[5pt]
    &= \frac{1}{N}g(f). \label{eq:dt_over_df_2}
\end{align}

By equating \eqref{eq:dt_over_df_1} and \eqref{eq:dt_over_df_2}, we correlate the given function $g(\cdot)$ with the probability density function $p(\cdot)$:
\begin{equation}
    \frac{1}{Z}p(f) = \frac{1}{N}g(f). \label{eq:p_and_g}
\end{equation}

We can determine $N$ by integrating both sides of \eqref{eq:p_and_g}:
\begin{equation}
	N = \int_{f_{\min}}^{f_{\max}} g(x) dx. \label{eq:sol_n}
\end{equation}

Using \eqref{eq:t_of_f} and \eqref{eq:p_and_g}, $u(f)$ is calculated from the known function $g(\cdot)$:
\begin{equation}
	u(f) = \frac{1}{N} \int_{f_{\min}}^{f} g(\tau) d\tau. \label{eq:sol_t_of_f}
\end{equation}

To retrieve $f(u)$, numerically invert $u(f)$ within the bounded range $u \in (0, 1)$. This procedure is summarized in Algorithm \ref{alg:diffeq}.

%% file: sections/simul-radial-isotropic.tex
\section*{Appendix 2: Isotropic 3D Radial Trajectory}

We consider a 3D Radial trajectory with an isotropic resolution and FOV:
\begin{align*}
    K(\phi) &= K = 1.25~\text{cm}^{-1},\\
    L(\phi) &= L = 10~\text{cm}.
\end{align*}
In this scenario, the trajectory can be constructed by locating $K$-points uniformly distributed on a sphere. Our goal is to verify whether the proposed method yields the identical solution as that of the conventional approach.

Given the specified resolution and FOV, the spacings in the latitude and longitude directions are the reciprocal of the FOV:
\begin{align}
    K \sin \phi \Delta \theta &= L^{-1},\\
    K \Delta \phi &= L^{-1}.
\end{align}

\noindent Consequently, the elemental area on the sphere becomes:
\begin{align}
    dA &= (K \sin \phi \Delta \theta) \cdot (K \Delta \phi) \nonumber\\
    &= L^{-2}.
\end{align}

\noindent With $K$-points uniformly distributed, the number of readouts is calculated by dividing the sphere's total surface area by the elemental area \cite{bernstein2004handbook}:
\begin{align}
    N &= \frac{A}{dA} \nonumber\\
    &= 4 \pi K^2 L^2 \label{eq:3dpr_isotropic_nrd_1}\\
    &\approx 1964. \nonumber
\end{align}

With the proposed method, the number of readouts is computed using \eqref{eq:3dpr_nrd}:
\begin{align}
    N &= 2 \pi L \int_0^\pi L K \sin \phi ~ d\phi \nonumber\\
    &= 4 \pi K^2 L^2 \label{eq:3dpr_isotropic_nrd_2}\\
    &\approx 1964. \nonumber
\end{align}
The equivalence between \eqref{eq:3dpr_isotropic_nrd_1} and \eqref{eq:3dpr_isotropic_nrd_2} confirms that the proposed method yields the same number of readouts as the conventional approach.

Next, we examine whether our method produces uniformly distributed samples on the sphere. From \eqref{eq:3dpr_u_of_phi}, the differential equation for $\phi_p(u)$ is:
\begin{align}
    \dot{\phi}_p(u) &= \frac{4 \pi K^2 L^2}{2 \pi K^2 L^2 \sin \phi_p(u)}\\
    &= \frac{2}{\sin \phi_p(u)},
\end{align}
\noindent leading to the closed-form solution:
\begin{equation}
    \phi_p(u) = \cos^{-1}(1 - 2u).
\end{equation}

\noindent For $\theta_p(u)$, the differential equation is:
\begin{align}
    \dot{\theta}_p(u) &= \frac{2 \pi K L}{\sqrt{u(1-u)}}.
\end{align}

\noindent Using Mathematica \cite{Mathematica}, the closed-form solution for $\theta_p(u)$ is derived:
\begin{equation}
    \theta_p(u) = (2 \pi K L) (-\pi - 2 j \log(j \sqrt{1-x} - \sqrt{x})).
\end{equation}

\begin{figure}[!htb]
    \centering
    \begin{minipage}{0.22\textwidth}
        \centering
        \includegraphics[width=0.89\linewidth]{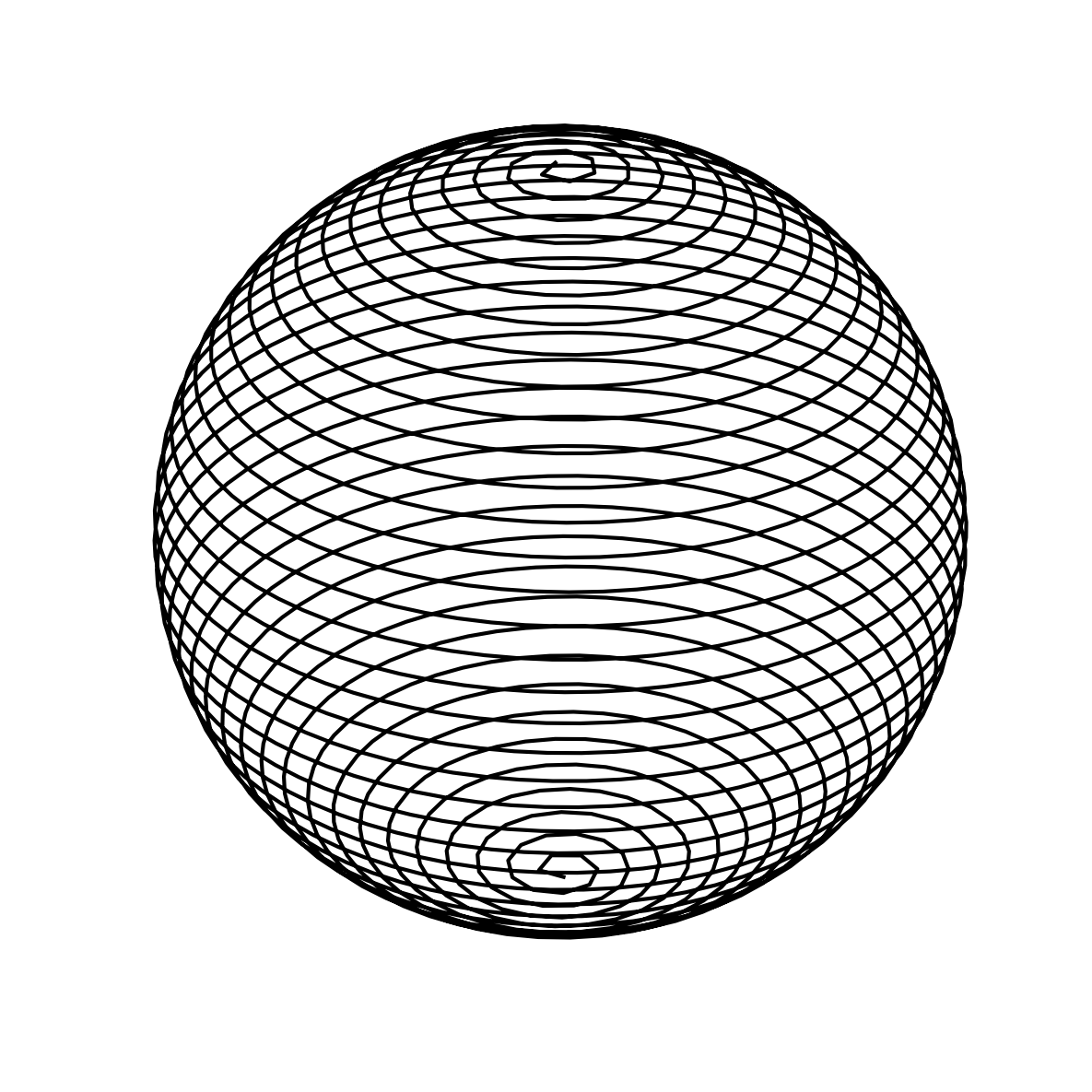}
        \caption*{(a) Spiral path on a sphere}
        \label{fig:3dpr_isotropic_spiral_path}
    \end{minipage}
    \begin{minipage}{0.22\textwidth}
        \centering
        \includegraphics[width=\linewidth]{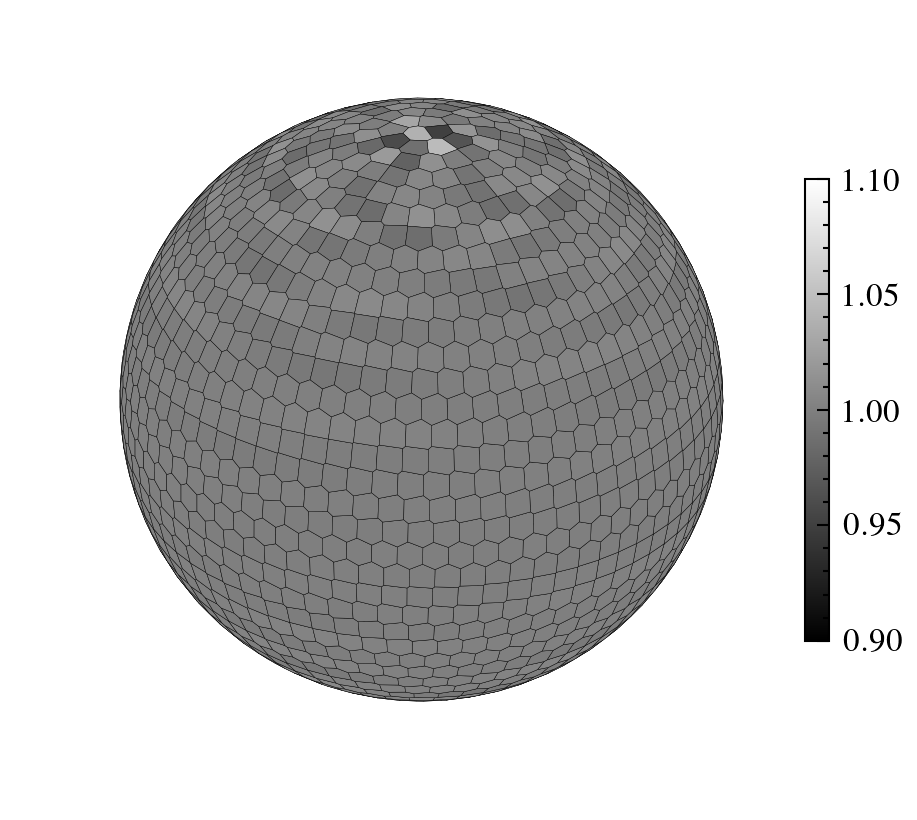}
        \caption*{(b) Voronoi regions}
        \label{fig:3dpr_isotropic_voronoi}
    \end{minipage}
    \caption[Spiral path and Voronoi regions on a sphere]{Spiral path and Voronoi regions on a sphere for isotropic FOV and resolution. Though there are a few exceptions near the pole, most of the samples are uniformly spread across the sphere's surface.}
    \label{fig:3dpr_isotropic}
\end{figure}

Figure \ref{fig:3dpr_isotropic} displays the resulting spiral path and the areas of Voronoi regions \cite{caroli2010robust}. Though there are a few exceptions near the pole, most of the samples are uniformly spread across the sphere's surface.

%% file: sections/simul-radial-anisotropic.tex
\section*{Appendix 3: Anisotropic 3D Radial Trajectory}

\begin{figure}[!htb]
    \centering
    \includegraphics[width=0.47\textwidth, keepaspectratio]{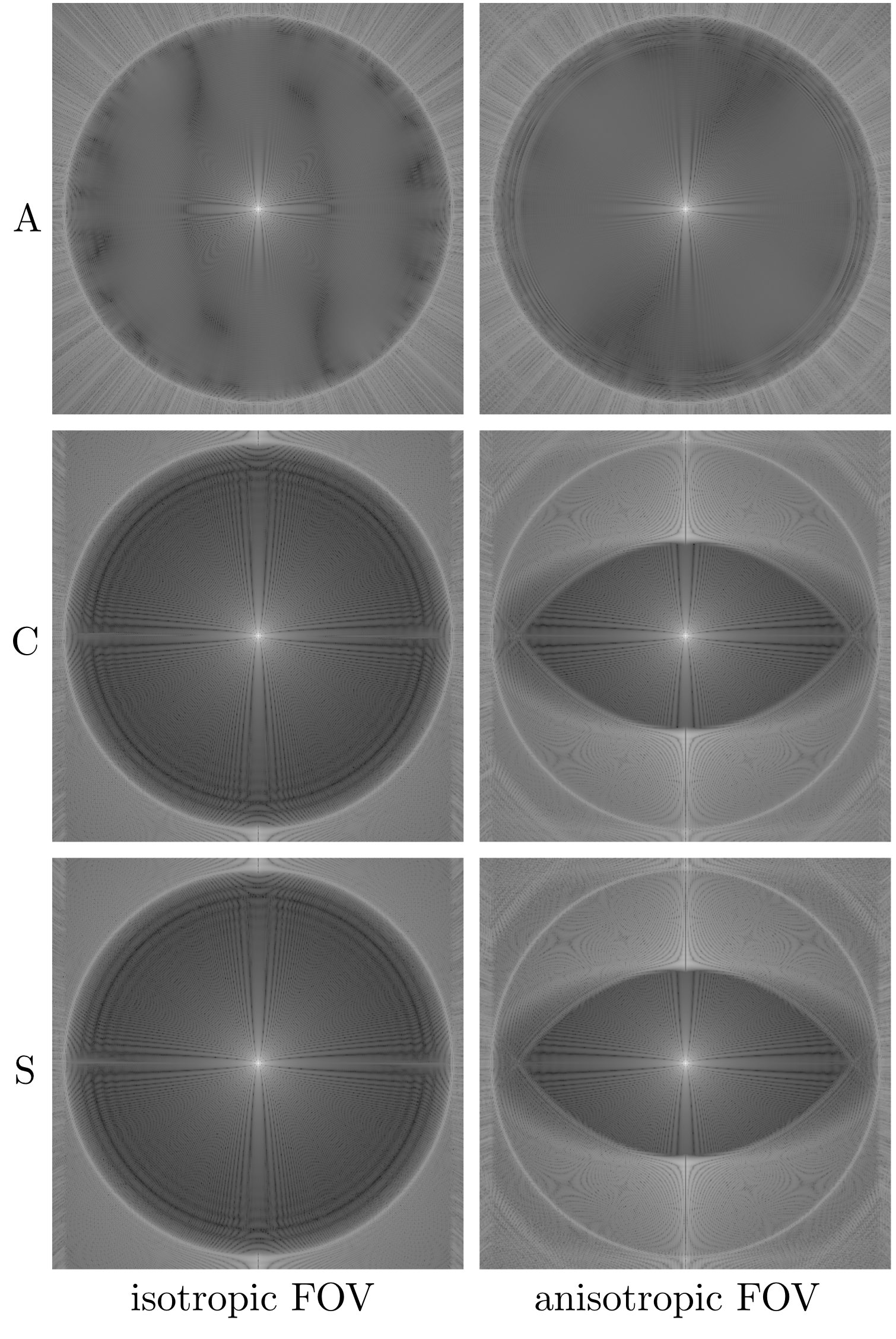}
    \caption[PSF of 3D Radial trajectories for isotropic and anisotropic FOVs]{PSFs of 3D Radial trajectories for an isotropic FOV of (28, 28, 28)-cm (left) and an anisotropic FOV of (28, 28, 14)-cm (right). From top to bottom: axial (A), coronal (C), sagittal (S) planes. The images are depicted over (60, 60)-cm areas. The PSFs confirm that the proposed method successfully produces the 3D Radial trajectories for the given imaging parameters.}
    \label{fig:traj-3dpr-psf}
\end{figure}

3D Radial trajectories were designed using the following imaging parameters: isotropic resolution of 1.2mm, maximum gradient amplitude of 39 mT/m, and maximum slew rate of 142 mT/m/ms. Figure~\ref{fig:traj-3dpr-psf} compares the point spread function (PSF) of 3D Radial trajectories with an isotropic FOV of (28, 28, 28)-cm and an anisotropic FOV of (28, 28, 14)-cm. The numbers of readouts for these trajectories were 171,042.27 and 103,412.12, respectively. The measured PSFs confirm that the proposed method successfully produces the 3D Radial trajectories for the given imaging parameters.

